\documentclass[12pt]{article}
\usepackage[utf8]{inputenc}
\usepackage{amsmath, amsthm, mathtools, amsfonts, graphicx, hyperref,setspace, xparse, algorithm, algorithmicx, algpseudocode, mathabx, physics, tikz, orcidlink}

\usepackage{algorithm}
\usepackage{algpseudocode}

\usepackage[margin=1in]{geometry}

\usepackage{biblatex, tikz}
\theoremstyle{plain}
\newtheorem{theorem}{Theorem}[section]
\newtheorem{lemma}[theorem]{Lemma}
\newtheorem{prop}[theorem]{Proposition}

\theoremstyle{definition}
\newtheorem{definition}[theorem]{Definition}
\newtheorem{example}[theorem]{Example}
\newtheorem{numexmp}[theorem]{Numerical Example}
\newtheorem{remark}[theorem]{Remark}

\title{Common random number simulation as the optimal coordinatewise coupling for iterated random functions}
\author{Sabrina Sixta\thanks{email: sabrina.sixta@mail.utoronto.ca} and Jeffrey S. Rosenthal\thanks{email: jeff@math.toronto.edu} \\
Department of Statistical Sciences, University of Toronto, Toronto, ON, 
	Canada}

\def\vecx{\vec{X}}

\newcommand{\indep}{\perp \!\!\! \perp}
\def\tin{\tau^{-1}}

\def\tinv{\tau^{-1}}

\def\unif{\mathrm{Unif}}
\def\bernoulli{\mathrm{Bernoulli}}
\def\d{\mathrm{d}}

\begin{document}
	\maketitle
 \begin{abstract}
 	In this paper, we extend the results of \cite{crn} to demonstrate that common random number (CRN) simulation applied to coordinatewise monotone iterated random functions (IRF) represents the optimal allocation of random numbers to minimize the expected distance of an $n$-step Markov chain (MC) in $\mathbb{R}^q$. We introduce this concept as the optimal coordinatewise Wasserstein distance (CWD) coupling. We provide specific examples of IRFs where CRN is the optimal CWD coupling, as well as cases where it is not. For IRFs that are not coordinatewise monotone, we illustrate how CRN applied to equimeasurable transformations can result in even lower expected distances compared to CRN applied to the original IRF.
 	
 \end{abstract}	
 
 
 \begin{tabbing}
 	 \> Secondary 49Q22, 60J20
 \end{tabbing}

	\tableofcontents
	\addtocontents{toc}{\protect\setcounter{tocdepth}{2}}
	
 
 

\section{Introduction}

Markov chain Monte Carlo (MCMC) is an integral part of statistical computing and Bayesian statistics is its most notable application \cite{honestexp,handbook}. An important question when simulating MCMC is how quickly the chain converges to the stationary distribution \cite{honestexp}. The Wasserstein distance is a metric on probability distributions that is a popular method for evaluating whether a chain is close to stationarity \cite{wasmeth, gibbswass}. It scales well in high dimensions compared to drift and minorization generated total variation bounds \cite{wasmeth} and metrizes convergence in distribution in $P_k(\mathcal{X})$ \cite{villani}.

Various methods for obtaining theoretical upper bounds on the Wasserstein distance between the distribution of a Markov chain (MC) and its stationary distribution have been developed \cite{wasmeth, wassdnm, localcont, waslinearmixedmodels}. However, such bounds can involve complex calculations, be difficult to implement, and are unavailable for many MCMC models of interest. An alternative method for estimating upper bounds on the Wasserstein distance is through common random number (CRN) simulation \cite{crn, biswas, sixtacrn} alternatively called synchronous coupling \cite{sdesynch,browniansynch}. 


We consider CRN in the setting of iterated random functions (IRFs). A MC $\{X_n\}_{n\geq 0}$ can be represented as a sequence of IRFs \cite{iteratedranfunc,localcont,avcont,ifssurvey,ifssurvey1}, where
$$
X_n=f(\theta_n,X_{n-1})
$$
and $\{\theta_n\}_{n\geq1}$ is a sequence of i.i.d. random variables. Consider two copies of the chain, $\{X_n\}_{n\geq0}$ and $\{Y_n\}_{n\geq0}$, initialized at $x_0$ and $y_0$, respectively. Under CRN, the same realization of $\theta_n$ is used to update both chains:
$$
X_n=f(\theta_n,X_{n-1}),
\qquad
Y_n=f(\theta_n,Y_{n-1}).
$$
Thus, CRN induces a coupling between the two copies of the chain.





The optimality of CRN as the joint distribution that minimizes the expected distance has been established in several settings. As early as 1934, it was shown that if $\theta, \theta'\sim Unif(0,1)$ and $f_X$ and $f_{Y}$ are quantile functions, then CRN simulation minimizes the Euclidean distance \cite{inequalities}.  Proposition 2.2 of \cite{crn} extends this result to coordinatewise increasing transforms. Related results have been established for stochastic processes and specific MCs. For example, \cite{browniansynch} shows that, under suitable conditions, two reflected Brownian motion processes simulated using CRN converge to one another almost surely. \cite{sdesynch} shows that CRN generates the optimal Wasserstein coupling between the solutions of stochastic differential equations.  \cite{independentmetropolishastings} shows that CRN generates a maximal coupling with respect to total variation distance for the independent Metropolis algorithm. However, the conditions under which CRN induces an optimal coupling that minimizes the expected distance between two copies of a Markov chain represented as an iterated random function have not been fully characterized.

In this paper we provide sufficient conditions for when CRN applied to an IRF minimizes the expected distance between two copies of an $n$-step MC in $\mathbb{R}^q$ (see Theorem \ref{thm:crnmc}). In particular, we generalize the results of \cite{crn} to quasi-antitone distance functions and extend CRN optimality results from single-step transformations to iterated random functions. We show that CRN could be the optimal coupling for a 1-step IRF, but not a 2-step IRF (See Lemma \ref{lem:ranlogmap} compared to Numerical Example \ref{numex:ranlogmap}). When our sufficient conditions are not satisfied, we demonstrate how alternative couplings can yield smaller expected distances (Numerical Example \ref{numex:ranlogmap}). 

In Section \ref{sec:invdf}, we summarize the useful properties of CRN when the random transform is the quantile function (ie., the inverse distribution function). We justify why the Wasserstein distance is the natural measure for studying the convergence properties of random values simulated using CRN. We also introduce Propostion \ref{prop:interval} (Theorem 2 of \cite{cambanis1976}), which lays the foundations for the rest of the theorems in the paper.


In Section \ref{sec:pwd}, we first show in Theorem \ref{thm:wasseqinf} that the Wasserstein distance between the laws of $\vec{X}=f(\vec{\theta},\vec{x})$ and $\vec{Y}=f(\vec{\theta}',\vec{y})$ is attained by minimizing with respect to the joint distribution between $\vec{\theta}$ and $\vec{\theta}'$. Next, we introduce the concept of coordinatewise Wasserstein distance (CWD) for IRFs, Definition \ref{def:pwd}. If we write the MC as an IRF, the CWD is the infimum expected distance between $f(\vec{\theta},\vec{x})$ and $f(\vec{\theta}',\vec{y})$ with respect to a product measure between $\vec{\theta}$ and $\vec{\theta}'$, $\bigotimes_{i=1}^d \mathcal{L}(\theta_i,\theta_i')$. We also provide an example of when the Wasserstein distance is strictly smaller than the CWD (Example \ref{ex:bernoulli}).

In Section \ref{sec:1step}, we identify when CRN is an optimal CWD coupling on the first iteration of a MC. In Theorem \ref{thm:rdrp} we show that if $f(\vec{\theta},\vec{x})$ is increasing coordinatewise over $\vec{\theta}$, then CRN on the first iteration of a MC is an optimal CWD coupling.

In Section \ref{sec:nstep} we extend the results of Section \ref{sec:1step} to provide sufficient conditions for when CRN is the optimal CWD coupling on the $n$th iteration of a MC. In Theorem \ref{thm:crnmc} we show that if $f(\vec{\theta},\vec{x})$ is increasing coordinatewise with respect to both $ \vec{\theta}$ and $\vec{x}$, then CRN on the $n$th iteration of a MC is an optimal CWD coupling.

In Section \ref{sec:ex}, we present a range of examples illustrating when CRN is, and is not, an optimal CWD coupling. We provide examples where CRN is the optimal CWD coupling for a 1-step chain and an $n$-step chain (see Section \ref{sec:ex}). We show that CRN could be the optimal CWD coupling for a 1-step chain, but not an $n$-step chain (see Example \ref{ex:randomlogmap} compared to Numerical Example \ref{numex:ranlogmap}). 

In Section \ref{sec:nonmon}, we look at ways to lower the expected distance when CRN is not an optimal CWD coupling. We propose generating an equimeasurable function that is monotone in $\vec{\theta}$, which will generate the optimal CWD coupling when CRN is applied. We provide two methods of generating equimeasurable transforms (see Lemmas \ref{lem:intflip} and \ref{lem:intmerge}) and apply our lemmas to an example.


First, we introduce notation in the Section \ref{sec:notationanddef}. This section should be referenced if the reader wants clarity on notation throughout the paper.

The code used to generate all of the tables can be found at \href{https://github.com/sixter/OptimalityOfCRN}{github.com/sixter/OptimalityOfCRN}.
		
\section{Background}\label{sec:background}

\subsection{Notation and definitions}\label{sec:notationanddef}
\paragraph{Markov chain notation}Let $\{X_n\}_{n\geq 1}$ be a MC such that $X_n$ can be written as an iterative random function ie. $X_n=f(\vec{\theta}_n,X_{n-1})\in \mathbb{R}$ and $\vec{\theta}_n$ is a random variable. 
Define $\Theta_n=(\vec{\theta}_1,\ldots, \vec{\theta}_n)$ where $\vec{\theta}_1,\ldots, \vec{\theta}_n$ are i.i.d. random variables. $X_n$ can then be written as follows for initial value $x_0$.
\begin{equation}\label{eq:forward}
 X_n=f(\vec{\theta}_n,f(\vec{\theta}_{n-1},\ldots, f(\vec{\theta}_1,x_0)\ldots ))=f_n(\Theta_n,x_0)
\end{equation} 
Any time-homogeneous MC can be represented as an IRF \cite{avcont}. $X_n$ is the state of the MC at iteration $n$. 

$\mathcal{L}(X_n)$ denotes the law of $X_n$ and $\mathcal{L}(X_n)\Rightarrow\pi$ denotes weak convergence of the law of $X_n$ to a stationary distribution $\pi$. We write that $x\in \text{supp}(\pi)$ if $x$ is in the support of $\pi$.

More generally, an $\mathbb{R}^q$-valued Markov chain $\{\vec{X}_n\}_{n\geq 1}$ can be represented as
$\vec{X}_n=f(\vec{\theta}_n,\vec{X}_{n-1})\in\mathbb{R}^q$ where, for each coordinate $l\in{1,\ldots,q}$, $X_n^{(l)}=f^{(l)}(\vec{\theta}_n,\vec{X}_{n-1})$.

 
 
\paragraph{Pushforward measure}
For a fixed $\vec{x}$, let $f_{\vec{x}}(\cdot):=f(\cdot,\vec{x})$. The pushforward of $\mu$ under $f_{\vec{x}}$ is denoted by $(f_{\vec{x}})_*\mu:\mathcal{B}(\mathbb{R}^q)\to[0,1]$. For any $B\in \mathcal{B}(\mathbb{R}^q)$, $(f_{\vec{x}})_*\mu(B):= \mu[(f_{\vec{x}})^{-1}(B)]$. 
Since $\vec{\theta}_{n}\sim \mu$ and $\vec{X}_{n}=f(\vec{\theta}_n, \vec{X}_{n-1})$, $\mathcal{L}(\vec{X}_n\mid\vec{X}_{n-1}=\vec{x})= (f_{\vec{x}})_{*}\mu$. See Section 2.4 of \cite{bridgeford2025measure} for more details on pushforward measures.

 
 \paragraph{CRN}
CRN simulation occurs when two Markov chains $\{X_n\}_{n\geq 1}$ and $\{Y_n\}_{n\geq 1}$, with different initial values $x_0\neq y_0$, are simulated using the same IRF and the same realizations of the random inputs. That is,
$$ X_n=f(\vec{\theta}_n,X_{n-1})
 \qquad\text{and}\qquad
 Y_n=f(\vec{\theta}_n,Y_{n-1}),\qquad n\geq 1,
$$
 use the same $\vec{\theta}_n$ at each iteration. Equivalently,
$$
 X_n=f_n(\Theta_n,x_0)
 \qquad\text{and}\qquad
 Y_n=f_n(\Theta_n,y_0),\qquad n\geq 1,
$$
 where the same $\Theta_n$ is used for both chains. The CRN coupling depends on the choice of IRF representation and is therefore not uniquely determined by the Markov chain \cite{crn}. Some treatments restrict CRN to the inverse distribution function (quantile) representation \cite{crn,jacob}, as considered in Section \ref{sec:invdf}.
 
\paragraph{Distance function $k$:}Let $k$ be a distance function. 

$k$ is quasi-antitone (in other words $-k$ is supermodular) if for all $x\leq x'$, $y\leq y'$,
$$k(x,y)+k(x',y')-k(x,y')-k(x',y)\leq 0$$
We will need $k$ to be quasi-antitone to prove that CRN is an optimal coupling. We write that $k$ is QAC if $k$ is a quasi-antitone continuous distance function. 

 The distance function corresponding to the $\ell^p$ norm $p\geq 1$, $||\vec{x}-\vec{y}||_p=\left(\sum_{i=1}^q |x_i-y_i|^p\right)^{1/p}$ is quasi-antitone with respect to each pair of coordinates.
 Theorem 3 of \cite{rearrangefunc} establishes that continuous functions $h(x-y)$ where $h$ is convex and $h(0)=0$ are QAC. Thus, $h(x_i- y_i)=|x_i-y_i|^p$ is QAC. 

\paragraph{Wasserstein distance} We define $\mathcal{L}(X,Y)$ as a joint distribution of random variables $X,Y$ and $\mathcal{J}(\mu,\nu)=\{\mathcal{L}(X,Y)\mid \mathcal{L}(X)=\mu, \mathcal{L}(Y)=\nu \}$ to be the set of all joint distribution functions with marginals $\mu,\nu$. 
The Wasserstein distance of order $p$ between distributions $\mu,\nu$ with distance function $k$ is defined as 
$W_{k,p}(\mu,\nu)=\inf_{\gamma\in \mathcal{J}(\mu,\nu)}E_{\gamma}[k(X,Y)^p]^{1/p}$ for $p\in [1,\infty)$. If $p=1$, then we denote the Wasserstein distance as $W_{k}(\mu,\nu)$. 
The joint distribution $\gamma_{opt}\in \mathcal{J}(\mu,\nu)$ is an optimal Wasserstein distance coupling if $E_{\gamma_{opt}}[k(X,Y)]=W_{k}(\mu,\nu)$ \cite{villani}. 


We define the Wasserstein space $P_k(\mathcal{X})$ as the space of Borel probability measures $\mu$ on the metric space $(\mathcal{X},k)$ such that
$\int_{\mathcal{X}} k(x_0,x),d\mu(x)<\infty$
for some $x_0\in\mathcal{X}$.

Define a sequence of random variables $\{X_n\}_{n\geq 1}$ in a Polish metric space $(\mathcal{X},k)$ with distributions in $\mathcal{L}(X_n)\in P_k(\mathcal{X})$. Define $\mathcal{L}(X)=\pi\in P_k(\mathcal{X})$. $X_n$ converges weakly to $\pi$ in $P_k(\mathcal{X})$ (denoted as $\mathcal{L}(X_n)\Rightarrow \pi$ in $P_k(\mathcal{X})$) if (1) $\mathcal{L}(X_n)\Rightarrow \pi$ and (2) for some $x_0\in \mathcal{X}$, $E[k(x_0,X_n)]\to E[k(x_0,X)]$ (see alternative conditions in Definition 6.8 of \cite{villani}).

\subsection{Applying CRN to quantile functions}\label{sec:invdf}
We first discuss the simplest case of CRN: when random variables $X,Y\in\mathbb{R}$ are written as the quantile transform of a uniform random variable. We define the quantile function as $F^{-1}(u)=\inf\{x\mid u\leq F(x)\}$, and our random variables as functions of the quantile function, $X=F^{-1}_{\mu}(U)\sim\mu$ and $Y=F^{-1}_{\nu}(V)\sim\nu$ where $U,V\sim Unif(0,1)$. Proposition  \ref{prop:interval} specifies the range of possible expected distances $E[k(X,Y)]$ over all elements in $\mathcal{J}(\mu,\nu)$.

Proposition \ref{prop:interval} lays the foundation for the rest of the theorems in this paper. Namely Theorem \ref{prop:incr1} uses Proposition \ref{prop:interval} to prove that CRN is the optimal coupling for the Wasserstein distance when $X$ is an increasing function of $U$. 

\begin{prop}[Theorem 2 of \cite{cambanis1976}]\label{prop:interval}
		Suppose that $X,Y \in \mathbb{R}$ are two random variables with corresponding distribution functions $F_{\mu}, F_{\nu}$ such that $X\sim \mu$ and $Y\sim \nu$. Suppose that $k$ is a QAC metric function and that $\mu,\nu\in P_k(\mathbb{R})$. 
		Then the set of possible expected distances over all couplings in \(\mathcal J(\mu,\nu)\) is the following interval. 
		$$\big\{E[k(X,Y)] \mid \mathcal{L}(X,Y)\in \mathcal{J}(\mu,\nu) \big\}= \big[E[k(F^{-1}_{\mu}(U),F^{-1}_{\nu}(U))], E[k(F^{-1}_{\mu}(U),F^{-1}_{\nu}(1-U))] \big]$$
		\begin{proof}
		The result follows from Theorem 2 of \cite{cambanis1976}. Note that condition $(a)$ of Theorem 2 of \cite{cambanis1976} is satisfied, because $k$ is symmetric and $\mu,\nu\in P_k(\mathbb{R})$, which means that for some $x_0\in\mathbb{R}$, $E[k(X,Y)]\leq E[k(X,x_0)]+E[k(x_0,Y)]<\infty$. 	
		\end{proof}	
\end{prop}	

Propositions \ref{thm:equivinv} and \ref{thm:equivinv4}, which are a combination of established results, summarize why we focus on Wasserstein distance to study convergence of CRN applied to quantile functions. Namely, 
 that CRN applied to the quantile function metrizes convergence in distribution in $P_k(\mathbb{R})$. 

The importance of this observation lies in the fact that it identifies the Wasserstein distance as a natural metric for studying convergence of random variables generated via CRN. This is in contrast to the total variation distance, which is the metric used to analyze CRN convergence in \cite{crnconv,independentmetropolishastings}. Total variation distance is strictly stronger than convergence in distribution (Theorem 6 of \cite{gibbs}). Indeed, convergence in total variation distance requires that $P(X=Y)\to 1$, which is equivalent to the existence of a coupling under which the two random variables become equal with probability tending to one  \cite{probsurv}. In most applications of CRN for real-valued random variables, however, the coupled random variables are not equal with probability tending to 1; rather, they are constructed to be strongly positively dependent.
\begin{prop}\label{thm:equivinv}
	Define the random variables $\{X_n\}_{n\geq 1}$ and $X$ in a Polish metric space $(\mathbb{R},k)$ such that $\sup_{n}E[k(X_n,x_0)]<\infty$.
	The following are equivalent where $U\sim \unif(0,1)$.
	\begin{enumerate}
		\item $\mathcal{L}(X_n)\Rightarrow \pi \text{ in } P_k(\mathbb{R})$
		\item $W_{k}(\mathcal{L}(X_n), \mathcal{L}(X))\to 0$
		\item $E[k(F^{-1}_{X_n}(U),F^{-1}_{X}(U))]\to 0$
	\end{enumerate}
	
	\begin{proof}
		
		$(1)\iff (2)$: by Theorem 6.9 of \cite{villani}.
		
		$(2)\iff (3)$: By Proposition \ref{prop:interval}, the Wasserstein distance is attained when CRN is applied for each $n$, $W_k(\mathcal{L}(X_n), \mathcal{L}(X))=E[k(F^{-1}_{X_n}(U),F^{-1}_{X}(U))]$.
	\end{proof}
\end{prop}

Proposition \ref{thm:equivinv4} shows that CRN applied to quantile functions can provide a stronger form of convergence than convergence in the Wasserstein distance. In particular, almost sure convergence of the CRN-coupled random variables follows from weak convergence of their distributions, without requiring convergence in $P_k(\mathbb{R})$. Thus, under the quantile coupling, weak convergence of the distributions can be strengthened to almost sure convergence of the coupled random variables.

Proposition \ref{thm:equivinv4} is closely related to Lemma 21.2 of \cite{vaart} and to the proof of Skorohod's representation theorem.
\begin{prop}\label{thm:equivinv4}
	Let $\{X_n\}_{n\geq 1}$ be random variables in the metric space $ (\mathbb{R},|\cdot|)$ and $\pi$ be a distribution. Then the following holds, 
	$$\mathcal{L}(X_n)\Rightarrow \pi \iff
	|F^{-1}_{n}(U)-F^{-1}(U)| \to 0 \quad\text{almost surely}$$
	
	Where $F_{n}$ and $F$ are the CDF of $\mathcal{L}(X_n)$ and $\pi$, respectively, and $U\sim Unif(0,1)$.  
	\begin{proof}
		See proof in Subsection \ref{pf:skorohod}, which closely follows the proof of Skorohod's theorem (see Theorem 25.6 in \cite{billingsley}). 
	\end{proof}
\end{prop}
The following example provides a case where $\mathcal{L}(X_n)\Rightarrow \pi$ in distribution but not in $P_k(\mathbb{R})$, so Proposition \ref{thm:equivinv} does not hold, but Proposition \ref{thm:equivinv4} holds. 

\begin{example}
	Let $\{X_n\}_{n\geq 1}$ be a sequence such that for each $n\geq 1$, $P(X_n=n)=\frac{1}{n}$ and $P(X_n=0)=1-\frac{1}{n}$. $\mathcal{L}(X_n)$ converges weakly to the Dirac mass at point 0, $\mathcal{L}(X_n)\Rightarrow \delta_0$. In this case $F_n$ and $F^{-1}_n$ are defined as follows,
	
	\noindent\begin{minipage}{.5\linewidth}
		\[ F_n(x)=\begin{cases} 
			0 & x<0 \\
			1-\frac{1}{n} & 0\leq x <n \\
			1 & n\leq x 
		\end{cases}
		\]
	\end{minipage}%
	\begin{minipage}{.5\linewidth}
			\[ F^{-1}_n(u)=\begin{cases} 
			0 & 0\leq u \leq 1-\frac{1}{n} \\
			n & 1-\frac{1}{n}< u \leq 1 
		\end{cases}
		\]
	\end{minipage}
	
	$\mathcal{L}(X_n)$ does not converge weakly to $\delta_0$ in $P_k(\mathbb{R})$: Fix a point $x_0$, $\lim_{n\to \infty} E[|x_0-X_n|]=\lim_{n\to\infty}|x_0-0|(1-\frac{1}{n})+|x_0-n|\frac{1}{n}=|x_0|+1 \neq |x_0-0|$. 
	
Proposition \ref{thm:equivinv4} applies: First note that $F_{\delta_0}^{-1}(u)=0$ for all $u\in (0,1]$. Fix $u\in(0,1)$, then for $n>\frac{1}{1-u}$, $F^{-1}_n(u)=0=F_{\delta_0}^{-1}(u)$. So $F_n^{-1}$ converges pointwise to $F_{\delta_0}^{-1}$ for all $u\in (0,1)$. Since $P(U\in(0,1))=1$, $|F^{-1}_n(U)-F^{-1}_{\delta_0}(U)|\to 0$ w.p. 1.
	
	Bullet (3) of Proposition \ref{thm:equivinv} does not apply: $E[|X_n-0|]=|n-0|\frac{1}{n}+|0-0|(1-\frac{1}{n})=1\not\to 0$.
\end{example}

While the results related to the convergence of inverse transforms are elegant, they are generally impractical for studying the optimality of CRN applied to MCs. We usually do not know the quantile function of $X_n$. Further, we can only apply inverse transforms when $X\in\mathbb{R}$ (see Section 2 of \cite{santambrogio}). In the next section we will formally define CWD, which will be used to provide sufficient conditions under which CRN is the optimal CWD coupling between two IRFs in $\mathbb{R}^q$. 

\section{Coordinatewise Wasserstein distance}\label{sec:pwd}

%
%
%
	

Let $\vec{X},\vec{Y}\in\mathbb{R}^q$ be random vectors represented as random functions as follows:
$$
\vec{X}=f(\vec{\theta},\vec{x}),
\qquad
\vec{Y}=f(\vec{\theta}',\vec{y}),
$$
Where $\vec{\theta},\vec{\theta}'$ are defined as in Definition \ref{def:theta}.

\begin{definition}[$\vec{\theta}$]\label{def:theta}
	Let $\mathcal{X}_i$, $i=1,\ldots,d$, be topological spaces, each equipped with a total order relation (ie. for any $\theta_i,\theta'_i\in \mathcal{X}_i$ either $\theta_i\leq \theta'_i$ or $\theta'_i\leq\theta_i$). Let $\vec{\theta}=(\theta_1,\ldots,\theta_d)$ be a random vector taking values in the product space $\prod_{i=1}^d\mathcal{X}_i$, with independent coordinates $\theta_i\sim\mu_i$. Then
	$
	\mathcal{L}(\vec{\theta})=\mu=\bigotimes_{i=1}^d\mu_i.
	$
\end{definition}
We first show that the infimum expected distance between $\vec{X}$ and $\vec{Y}$ over all couplings of their distributions is equal to the infimum expected distance between $f(\vec{\theta},\vec{x})$ and $f(\vec{\theta}',\vec{y})$ over all couplings of $\vec{\theta}$ and $\vec{\theta}'$. Specifically,
\begin{equation}\label{eq:2}
	\begin{aligned}
		\inf_{\gamma\in
			\mathcal{J}(\mathcal{L}(\vec{X}),\mathcal{L}(\vec{Y}))}
		E_{(\vec{X},\vec{Y})\sim \gamma}[k(\vec{X},\vec{Y})]
		&=
		\inf_{\gamma\in
			\mathcal{J}(\mu,\mu)}
		E_{(\vec{\theta},\vec{\theta}')\sim \gamma}\left[
		k\bigl(f(\vec{\theta},\vec{x}),
		f(\vec{\theta}',\vec{y})\bigr)
		\right].
	\end{aligned}
\end{equation}

The equality in Equation \ref{eq:2} is formalized in the following theorem.


%
%
%
\begin{theorem}\label{thm:wasseqinf}
	Suppose that $\vec{X}=f(\vec{\theta},\vec{x})$ and $  \vec{Y}=f(\vec{\theta}',\vec{y})\in \mathbb{R}^q$ where $\vec{\theta},\vec{\theta}'$ are defined in Definition \ref{def:theta} and $f(\cdot,\vec{x})$ and $f(\cdot,\vec{y})$ are measurable functions. Then,
	\begin{align}\label{eq:wasseqinf}
		W_k(\mathcal{L}(\vec{X}), \mathcal{L}(\vec{Y}))=\inf_{\gamma\in
			\mathcal{J}(\mu,\mu)}
		E_{(\vec{\theta},\vec{\theta}')\sim \gamma}[k(f(\vec{\theta},\vec{x}),f(\vec{\theta}',\vec{y}))]
	\end{align}
	
	The proof is in Section \ref{pf:wasseqinf}
\end{theorem}

However, similar to \cite{crn}, the infima in the theorems of this paper are not taken over all couplings in $\mathcal{J}(\mu,\mu)$, so Theorem \ref{thm:wasseqinf} does not apply. Instead, we restrict the infimum to couplings in $\mathcal{J}_{IND}(\mu,\mu)\subset\mathcal{J}(\mu,\mu)$, which allow dependence only between corresponding components of $\vec{\theta}$ and $\vec{\theta}'$.


\begin{align*}
	\mathcal{J}_{IND}(\mu,\mu)&= \left\{\gamma \in	P(\mathcal{X}^d \times \mathcal{X}^d) \mid \gamma_i\in \mathcal{J}(\mu_i,\mu_i)\text{ and } \gamma =\bigotimes_{i=1}^d\gamma_i
\right\}
\end{align*}

This class is denoted by $\mathcal{M}_0(\mu,\mu)$ in \cite{crn}. The infimum in Proposition 2.2 of \cite{crn} and in Theorems \ref{prop:rd}, \ref{thm:rdrp}, and \ref{thm:crnmc} is taken over this restricted class of couplings.

Next, we introduce the coordinatewise Wasserstein distance, which is the infimum expected distance over all couplings in $\mathcal{J}_{IND}(\mu,\mu)$.
\begin{definition}[coordinatewise Wasserstein distance (CWD)]\label{def:pwd}
	The coordinatewise Wasserstein distance for IRFs is the infimum expected distance between $\vec{X} = f(\vec{\theta}, \vec{x})$ and $\vec{Y} = f(\vec{\theta}', \vec{y})$ over the set of distributions $ \mathcal{J}_{IND}(\mu,\mu)$.
	\begin{align*}
		\widetilde{W}_k(\mathcal{L}(\vec{X}), \mathcal{L}(\vec{Y})) &=	\inf_{\gamma\in
			\mathcal{J}_{IND}(\mu,\mu)}
		E_{(\vec{\theta},\vec{\theta}')\sim \gamma}[k(f(\vec{\theta},\vec{x}),f(\vec{\theta'},\vec{y}))]
	\end{align*}
\end{definition} 

Note that $\widetilde{W}_k(\mathcal{L}(\vec{X}), \mathcal{L}(\vec{Y}))\coloneqq \widetilde{W}_k((f_{\vec{x}})_*\mu, (f_{\vec{y}})_*\mu)$, with notation being used interchangeably throughout the paper.
$\gamma_{opt}\in \mathcal{J}_{IND}(\mu,\mu)$ is an optimal CWD coupling if $$E_{(\vec{\theta},\vec{\theta}')\sim\gamma_{opt}}[k(f(\vec{\theta},\vec{x}),f(\vec{\theta}',\vec{y}))]=\widetilde{W}_{k}((f_{\vec{x}})_*\mu,(f_{\vec{y}})_*\mu)$$


Since the CWD minimizes over $\mathcal{J}_{IND}(\mu,\mu)\subset \mathcal{J}(\mu,\mu)$, it is greater than or equal to the corresponding Wasserstein distance. That is,
\begin{align}\label{eq:glassyao}
	W_k(\mathcal{L}(\vec{X}), \mathcal{L}(\vec{Y}))\leq 
	\widetilde{W}_k(\mathcal{L}(\vec{X}), \mathcal{L}(\vec{Y}))
\end{align}
However, equality can hold under additional structural conditions. In particular, \cite{crn} establishes equality with respect to the Euclidean distance cost for certain queueing systems that satisfy coordinatewise monotonicity and permutability.

If $d=1$ in Definition \ref{def:theta}, Equation \ref{eq:glassyao} becomes an equality since $\mathcal{J}_{IND}(\mu,\mu)= \mathcal{J}(\mu,\mu)$. That is, 
\begin{align}\label{eq:pwdrequality}
	W_k(\mathcal{L}(X), \mathcal{L}(X'))=
	\widetilde{W}_k(\mathcal{L}(X), \mathcal{L}(X'))
\end{align}

Theorem 2 of \cite{Sagiv2019Wasserstein} provides an upper bound on the Wasserstein distance between push forward measures $\mathcal{L}(\vec{X})=(f_{\vec{x}})_*\mu$ and $\mathcal{L}(\vec{Y})=(f_{\vec{y}})_*\mu$ based on the same initial distributions $\mu$ and two different functions $f_{\vec{x}}, f_{\vec{y}}$. The inequality is easy to interpret, but the bound does not appear to be very tight for the random logistics map, which we suspect is the case for other examples as well. 
Theorem 3.2 of \cite{Carlier2025Quantitative} provides an upper bound on the Wasserstein distance of the pushforward measure based on the Wasserstein distance using optimal transport maps. 

In Example \ref{ex:bernoulli}, we provide a counterexample to show that the inequality in Equation \ref{eq:glassyao} can be strict if the joint couplings are restricted to $\mathcal{J}_{IND}(\mu,\mu)$.

\begin{example}\label{ex:bernoulli}
	Let $\theta_1,\theta_2\sim \text{Bernoulli}(\frac{1}{2})=\mu$ such that $\theta_1\indep\theta_2$. Denote $$f(\vec{\theta},x)=(\theta_1+2\theta_2)I_{x<2}+(\theta_2+2\theta_1)I_{x\geq 2}$$ 
	So, $f(\vec{\theta},1)=\theta_1+2\theta_2$ and $f(\vec{\theta}',2)=\theta'_2+2\theta'_1$. It is clear that $W_k((f_{1})_*\mu, (f_2)_*\mu)=0$ since $(f_1)_*\mu\coloneqq\mathcal{L}(f(\vec{\theta},1))=\mathcal{L}(f(\vec{\theta},2))\coloneqq (f_2)_*\mu$. 
	
	By Theorem \ref{prop:rd} and since $f(\cdot,x)$ is coordinatewise monotone and continuous in $\vec{\theta}$ for fixed $x\in \{1,2\}$, the CWD between the laws of $f(\vec{\theta},1)$ and $f(\vec{\theta},2)$ is attained when applying CRN,
	$$\widetilde{W}_k((f_1)_*\mu, (f_2)_*\mu) = 
	E[|f(\vec{\theta},1)-f(\vec{\theta},2)|]=E[|\theta_1-\theta_2|]=\frac{1}{2}$$
	
	This is an example of when the CWD ($\frac{1}{2}$) is strictly greater than the Wasserstein distance ($0$). 
	
	Finally, when I set $\theta_1=\theta'_2$ and $\theta_2=\theta'_1$, the Wasserstein distance is obtained,
	$E[|f((\theta_1,\theta_2),1)-f((\theta_2,\theta_1),2)|]=0=W_{|\cdot|}((f_{1})_*\mu, (f_2)_*\mu)$.
	Note that setting $\theta_1=\theta'_2$ and $\theta_2=\theta'_1$ is a coupling that is not defined in $\mathcal{J}_{IND}(\mu,\mu)$, but is defined in $\mathcal{J}(\mu,\mu)$.
\end{example}

\section{Conditions when CRN is an optimal coupling for MCs}\label{sec:crnopt}

In this section we first provide sufficient conditions for when CRN is an optimal coupling on the first iteration of a MC (Section \ref{sec:1step}) and then extend it to $n$ iterations of a MC (Section \ref{sec:nstep}).

	\subsection{Conditions when CRN is an optimal CWD coupling on the first iteration of a MC}\label{sec:1step}

In this section we provide sufficient conditions for when CRN is the optimal coupling on the first iteration. Recall the IRF notation where $\vec{X}_{n}=f(\vec{\theta}_n, \vec{X}_{n-1})\in \mathbb{R}^q$ is a MC and $\theta_n\in \prod_{i=1}^d \mathcal{X}_i$ is defined in Definition \ref{def:theta}. We first start with the simplest case when $d=q=1$ (Section \ref{subsec:rr}), extend this to the case when $d\geq 1$ and $q=1$ (Section \ref{subsec:rdr}), and finally generalize this to the case when $d,q\geq 1$ (Section \ref{subsec:rdrq}). In the simplest case we prove optimal Wasserstein distance coupling using Theorem \ref{thm:wasseqinf} while in the second and third cases we prove optimal CWD coupling.

\subsubsection{Conditions when CRN is an optimal Wasserstein distance  coupling for $d=q=1$}\label{subsec:rr}
The following theorem provides conditions under which the expected distance using CRN between two non-decreasing random functions in $\mathbb{R}$ is equal to the Wasserstein distance. This theorem is a generalization of Proposition 2.2 of \cite{crn}, which says that the covariance between two non-decreasing random functions  is maximized when generated by CRN. Before we specify the theorem we introduce the concept of monotonicity with respect to $\theta$. 

\begin{definition}[Monotone in $\theta$]\label{def:monotheta}
	Suppose that $f(\theta,x)$ is an IRF where $\theta\sim \mu$ and $x\in\mathbb{R}$. The function $f$ is considered to be monotone in $\theta$, if for all $\theta\leq \theta'$ $\mu$-a.e. and for all $x$ $\pi$-a.e.
	\begin{equation}
		f(\theta,x)\leq f(\theta',x)
	\end{equation} 
\end{definition}	
Theorem \ref{prop:incr1} states that if the functions $ f(\cdot,x)$ and $f(\cdot,y)$ are increasing and defined on the real line, then the expected distance using CRN equals the Wasserstein distance between $\mathcal{L}(X)$ and $\mathcal{L}(Y)$. 
\begin{theorem}
	\label{prop:incr1}
	
	Let $X$ be a random variable such that $\mathcal{L}(X)\in P_k(\mathbb{R})$ and $k$ is a QAC metric.  Suppose that $X=f(\theta,x)$ can be written as a random function where $\theta\sim \mu$ is a random variable. Define $Y=f(\theta',y)$ where  $\mathcal{L}(\theta')=\mathcal{L}(\theta)=\mu$ and $x,y\in \text{supp}(\pi)$.

	Suppose that $f$ is monotone and right-continuous in $\theta$. 
	Then CRN is the optimal Wasserstein distance coupling between $\mathcal{L}(X)$ and $\mathcal{L}(Y)$,
	\begin{equation}
		W_k(\mathcal{L}(X), \mathcal{L}(Y)) =E[k(f(\theta,x),f(\theta,y))]
	\end{equation}
	The proof is in Section \ref{pf:incr1}.
\end{theorem}

\begin{remark}[Connection to Brenier's Theorem]\label{rem:brenier}\label{prop:brenier}
Within the proof of Theorem \ref{prop:incr1}, we used Proposition \ref{prop:interval}, however, we could have also used Brenier's theorem, with specified $k(x,y)=(x-y)^2$, which coincides with Proposition 2.2 of \cite{crn}. The following is a simplified statement of Brenier's theorem in $\mathbb{R}$ (Theorem 2.1 of \cite{cuturi}, \cite{brenier}).
\begin{itemize}
	\item[] Suppose $X,Y\in \mathbb{R}$ are random variables, $k(x,y)=(x-y)^2$, and $\mathcal{L}(X),\mathcal{L}(Y)\in P_k(\mathbb{R})$. If $X$ has a density with respect to the Lebesgue measure, then the joint distribution that generates the optimal Wasserstein distance, $\mathcal{L}_{opt}(X, Y)$, can be defined as $X$ and $Y=T(X)$ where $T$ is the derivative of a convex function.
\end{itemize}

Let $X=g(\theta)$ and $Y=h(\theta')$ where $\mathcal{L}(\theta)=\mathcal{L}(\theta')$. Brenier's theorem implies that if $g$ and $h$ are increasing functions, then CRN is the optimal Wasserstein distance coupling:
First note that $k(x,y)=(x-y)^2$ is a QAC function. Next note that if $g$ is an increasing function, then $g^{-1}(x)=\inf\{u\mid g(u)\geq x\}$ is an increasing function. The coupling $\mathcal{L}_{opt}(X, Y)$ where $Y=h(g^{-1}(X))$ is optimal according to Brenier's theorem, because $T=h\cdot g^{-1}$ is an increasing function and thus is the derivative of a convex function. Now we remark that $g^{-1}(X)\sim\mathcal{L}(\theta)$ and so for $\theta\sim \mathcal{L}(\theta)$, $X=g(\theta)$ and $Y=h(\theta)$ thus implying that CRN is the optimal Wasserstein distance coupling. This is only a sketch, however, a more thorough explanation can be found in Section 2.1 of \cite{santambrogio} and Section 2.3.1 of \cite{jacob}.

If our MC is in $\mathbb{R}^q, q\geq 1$, the state space is no longer naturally endowed with a total order and so the definition of monotonicity in $\mathbb{R}$, Definition \ref{def:monotheta}, needs additional conditions.  Brenier specifies the optimal mapping to be the gradient of a convex function, which can be applied to MCs in $\mathbb{R}^q$. In \cite{crn} and in this paper the optimal coupling occurs when CRN is applied to functions that are coordinatewise increasing. 
\end{remark}

\subsubsection{Conditions when CRN is an optimal CWD coupling for $d\geq 1$ and $q=1$}\label{subsec:rdr}

By Definition \ref{def:theta} each $\theta_i\in \vec{\theta}$ has an order relation. Below, we define monotonicity over $\vec{\theta}$.

\begin{definition}[Monotone in $\theta_i$]\label{def:monovecthetai}
	Let $f(\vec{\theta},\vec{x})\in \mathbb{R}^q$ be an iterative random function where $\vec{x}\in \mathbb{R}^q$ and $\vec{\theta}$ is defined as in Definition \ref{def:theta}. $f$ is considered to be monotone in $\theta_i$, if for all $\theta_{i}\leq \theta'_{i}$ $\mu_i$-a.e. and $\vec{\theta}_{-i}$ $\mu_{-i}$-a.e.,  $\vec{x}$ $\pi$-a.e.
	\begin{equation}
		f^{(l)}((\theta_1,\ldots,\theta_{i},\ldots, \theta_d),\vec{x})\leq f^{(l)}((\theta_1,\ldots,\theta'_{i},\ldots, \theta_d),\vec{x})
	\end{equation} 
	for all $l\in \{1\ldots q\}$
\end{definition}	

\begin{definition}[Coordinatewise monotone in $\vec{\theta}$]\label{def:monovectheta}
The function $f$ is considered to be coordinatewise monotone in $\vec{\theta}$, if it is monotone in  $\theta_i$ for all $i\in\{1,\ldots,d\}$.
\end{definition}	

Next we will provide sufficient conditions for when CRN generates the optimal CWD coupling on the first iteration of a MC when $\vec{\theta}$ is defined as in Definition \ref{def:theta}. Similar to Theorem \ref{prop:incr1}, Theorem \ref{prop:rd} is a generalization of Proposition 2.2 of \cite{crn}.

\begin{theorem}
	\label{prop:rd}
	Let $X$ be a random variable
	such that $\mathcal{L}(X)\in P_k(\mathbb{R})$ and $k$ is a QAC metric. Suppose that $X=f(\vec{\theta},x)$ can be written as a random function where $\vec{\theta}$ is defined in Definition \ref{def:theta}. 
	Let $Y=f(\vec{\theta}',y)$ where  $\mathcal{L}(\vec{\theta}')=\mathcal{L}(\vec{\theta})=\mu$ and $x,y\in \text{supp}(\pi)$.
	
	Suppose further that $f$ is coordinatewise monotone and continuous in $\vec{\theta}$.
	Then the CWD between $\mathcal{L}(X)$ and $\mathcal{L}(Y)$ is attained when CRN is applied,
	\begin{align}\label{eq:rdconc}
		\widetilde{W}_k(\mathcal{L}(X), \mathcal{L}(Y))=
		E[k(f(\vec{\theta},x),f(\vec{\theta},y))]
	\end{align}
	The proof is in Section \ref{pf:rdinduction}.
\end{theorem}

\subsubsection{Conditions when CRN is an optimal CWD coupling for $d,q\geq 1$}\label{subsec:rdrq}
Finally, we generalize Theorem \ref{prop:rd} to apply to random variables that are defined in $\mathbb{R}^q, q\geq 1$. 

\begin{theorem}
	\label{thm:rdrp}
	Let $\vec{X}$ be a random vector such that $\mathcal{L}(\vec{X})\in P_{||\cdot||_p^p}(\mathbb{R}^q)$. Suppose that $\vec{X}=f(\vec{\theta},\vec{x})$ can be written as a random function where $\vec{\theta}$ is defined in Definition \ref{def:theta}. 
	Let $\vec{Y}=f(\vec{\theta}',\vec{y})$ where  $\mathcal{L}(\vec{\theta})=\mathcal{L}(\vec{\theta'})=\mu$ and $\vec{x},\vec{y}\in \text{supp}(\pi)$.
	
	Suppose further that $f$  is coordinatewise monotone and continuous in $\vec{\theta}$. 
	Then the CWD between $\mathcal{L}(\vec{X})$ and $\mathcal{L}(\vec{Y})$ is attained when CRN is applied,
	\begin{align}\label{eq:rdrpconc}
		\widetilde{W}_{||\cdot||_p,p}(\mathcal{L}(\vec{X}), \mathcal{L}(\vec{Y}))=
		E[||f(\vec{\theta},\vec{x})-f(\vec{\theta},\vec{y})||_p^p]^{1/p}
	\end{align}
	
	The proof is in Section \ref{pf:rdrp}.
\end{theorem}
	
	\subsection{Conditions when CRN is an optimal CWD coupling on the $n$th iteration of a MC}\label{sec:nstep}
	So far we've provided conditions for when CRN is the optimal CWD coupling on the first iteration of the MC. Next, in Theorem \ref{thm:crnmc}, we show that if a MC is coordinatewise monotone with respect to $\vec{\theta}$ and $\vec{x}$ then CRN is an optimal CWD coupling at any iteration of the MC.
	
	\begin{theorem}\label{thm:crnmc}
		
		Let $\{\vec{X}_n\}_{n\geq 1}$ be a MC such that $\mathcal{L}(\vec{X}_n)\in P_{||\cdot||_p^p}(\mathbb{R}^q), n\geq 1$. Suppose that $\vec{X}_n=f(\vec{\theta}_n,\vec{X}_{n-1})=f_n(\Theta_n,\vec{x}_0)$ can be written as a random function where where $\Theta_n=(\vec{\theta}_1,\vec{\theta}_2,\ldots, \vec{\theta}_n)$ are i.i.d. with distribution $\vec{\theta}_j\sim \mu =\bigotimes_{i=1}^d \mu_i$ and $\vec{x}_0\in \mathbb{R}^q$ is the initial value. Suppose that $f(\vec{\theta},\vec{x})$ is coordinatewise monotone and continuous in $\vec{\theta}$ and $\vec{x}$.
		
		Define $\vec{Y}_n=f(\vec{\theta}'_n,\vec{Y}_{n-1})=f_n(\Theta_n, \vec{y}_0)$ where  $\mathcal{L}(\vec{\theta}_n)=\mathcal{L}(\vec{\theta}'_n)=\mu$ and $\vec{x}_0,\vec{y}_0\in \text{supp}(\pi)$.
		Then the CWD between $\mathcal{L}(\vec{X}_n)$ and $\mathcal{L}(\vec{Y}_n)$ is attained when using CRN,
		\begin{equation}\label{eq:crnmc}
		\widetilde{W}_{||\cdot||_p,p}(\mathcal{L}(\vec{X}_n),\mathcal{L}(\vec{Y}_n))= 
			 E[||f_n(\Theta_n,\vec{x}_0)-f_n(\Theta_n,\vec{y}_0)||_p^p]^{1/p}
		\end{equation}
		The proof is in Section \ref{pf:crnmc}
	\end{theorem}
	\begin{remark}\label{rem:crnmc}
		If $X_n,Y_n\in \mathbb{R}$, then Theorem \ref{thm:crnmc} still applies for distance function $k$ QAC and Equation \ref{eq:crnmc} is replaced by the following.
		\begin{align}
			\widetilde{W}_{k}(\mathcal{L}(X_n),\mathcal{L}(Y_n))=
			E[k(f_n(\Theta_n,x_0),f_n(\Theta_n,y_0))]
		\end{align}
	\end{remark}

\section{Examples}\label{sec:ex}

The following section presents examples of MCs for which CRN yield an optimal CWD coupling. In the case of the random logistic map (Example \ref{ex:randomlogmap}), CRN applied to the one-step transition is optimal; however, for the $n$-step chain with $n\geq 2$, it is no longer optimal. This is demonstrated in Numerical Example \ref{numex:ranlogmap}, where we show that CRN fails to be the optimal CWD coupling for the two-step random logistic map. In contrast, for all other examples considered in this section, CRN provides an optimal CWD coupling for the $n$-step chain. 
\subsection{Random logistic maps}

\begin{example}[Random logistic maps]\label{ex:randomlogmap}
	Let $\{X_n\}_{n\geq 1}\in [0,1]$ be a random logistic map (\cite{localcont,wasandoneshot}). That is for i.i.d. $\beta_n\sim \text{Beta}(\alpha+1/2,\alpha-1/2)$ ($\alpha>1/2$), 
	$$X_n=f(\beta_n,X_{n-1})=4\beta_n X_{n-1}(1-X_{n-1})$$
\end{example}

In Lemma \ref{lem:ranlogmap}, we show that CRN is an optimal Wasserstein distance coupling on the first iteration of the random logistic map. Section \ref{subsec:ranlogmap2} illustrates why CRN applied to the 2-step random logistic map is sometimes not the optimal CWD coupling (depending on the initial values). 

\begin{lemma}\label{lem:ranlogmap}
	CRN is an optimal Wasserstein distance coupling on the first iteration of the random logistic map when $k$ is a QAC metric. 
	\begin{proof}
		Since $4X_{n-1}(1-X_{n-1})> 0$ w.p. 1, the function $4\beta_n X_{n-1}(1-X_{n-1})$ is increasing with respect to $\beta_n$. By Theorem \ref{prop:incr1}, CRN is an optimal Wasserstein distance coupling on the first iteration of the MC when $k$ is QAC. 
	\end{proof}
\end{lemma}

For all $\alpha>1/2$ the random logistic map converges in distribution to the stationary distribution,  $\text{Beta}(\alpha,\alpha)$. For $\alpha\leq 1$ the system is not locally contractive, and for $\alpha\geq2$ the system is locally contractive  (Proposition 7 of \cite{localcont}) and geometrically ergodic (Theorem 1 of \cite{localcont}). What is unknown is whether the expected distance between two MCs simulated using CRN converges to 0 when $\alpha\leq 1$.

\begin{figure}
	\centering
	\includegraphics[width=0.7\linewidth]{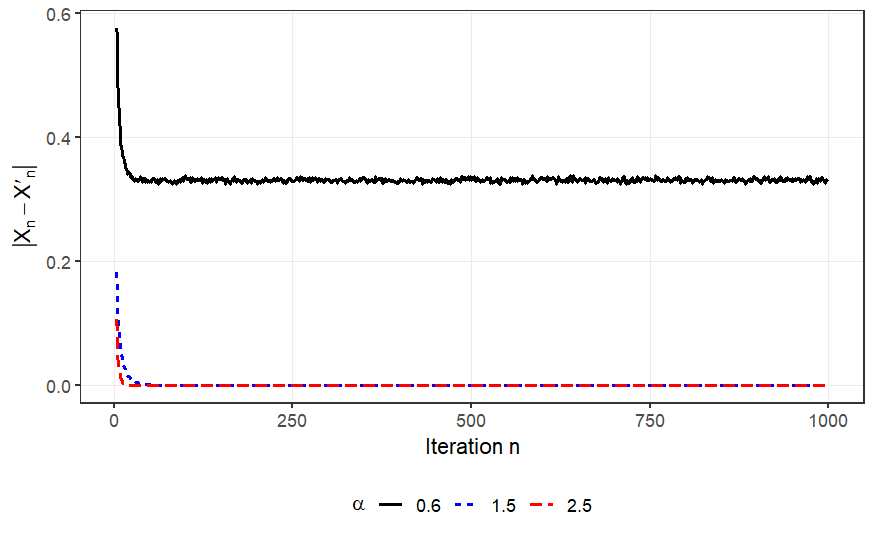}
	\caption{The simulated expected value of $|X_n-X'_n|$ where $X_n,X'_n$ are random logistic maps and CRN is applied ($X_0=0.5,X'_0=0.125$). For $\alpha>1/2$, the chain converges in distribution. For $\alpha\leq 1$ the chain is not locally contractive. For $\alpha> 2$, the chain is locally contractive. The CRN estimate uses the mean of $10 000$ simulations.}
	\label{fig:randomlogmapcrn}
\end{figure}

Figure \ref{fig:randomlogmapcrn} provides the simulated expected distance when CRN is applied for different $\alpha\in \{0.6,1.5,2.5\}$. The red line corresponding to $\alpha=2.5$ converges quickly to 0, which is expected, because it is geometrically ergodic and locally contractive. The blue line corresponding to $\alpha=1.5$ also converges quickly to 0. 

On the other hand, when $\alpha=0.6$, the simulated expected distance does not appear to converge to $0$ (at iteration $100\text{ }000$, $E[|X_{100\text{ }000}-X'_{100\text{ }000}|]\approx 0.343$, which is nearly unchanged from iteration $100$, $E[|X_{100}-X'_{100}|]\approx 0.331$). This indicates that when the random logistics map is not locally contractive, the two chains simulated using CRN may not converge.

\subsection{Zeno's walk}
\begin{example}[Zeno's walk]\label{ex:zeno}
	Let $\{X_n\}_{n\geq 0}\in \mathbb{R}$ be a Zeno's walk (\cite{localcont}). That is for i.i.d. $B_n\sim \bernoulli(r)$, 
	$X_n=B_n f^{+}(X_{n-1})+(1-B_n)f^{-}(X_{n-1})$ where,
	
	\noindent\begin{minipage}{.5\linewidth}
		\[ f^{+}(x)=\begin{cases} 
			x+1 & 0\leq x \\
			\frac{x}{2}+1 & -2\leq x\leq 0 \\
			x+2 & x\leq -2 
		\end{cases}
		\]
	\end{minipage}%
	\begin{minipage}{.5\linewidth}
		\[ f^{-}(x)=\begin{cases} 
			x-1 & x\leq 0 \\
			\frac{x}{2}-1 & 0\leq x\leq 2 \\
			x-2 & 2\leq x 
		\end{cases}
		\]
	\end{minipage}
\end{example}

\begin{figure}
	\centering
	\includegraphics[width=0.7\linewidth]{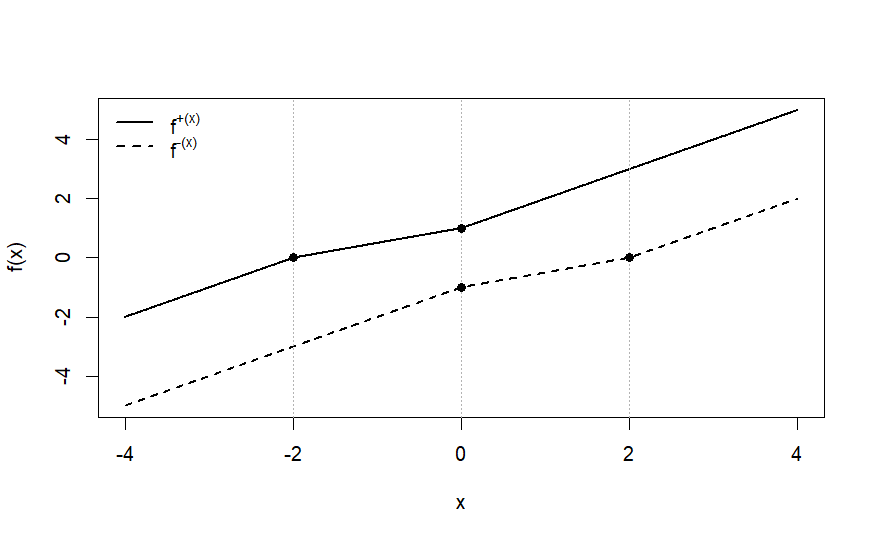}
	\caption{Graph of $f^{+}(x)$ and $f^{-}(x)$. It is clear from the graph that $f^{+}(x)-f^{-}(x)\geq 0$ for any $x\in \mathbb{R}$.}
	\label{fig:zenoswalk}
\end{figure}

See Figure \ref{fig:zenoswalk} for a graphical representation of $f^{+}(x)$ and $f^{-}(x)$.

Zeno's walk is locally contractive and geometrically ergodic if $\frac{1}{3}<r<\frac{2}{3}$ \cite{localcont}. In Lemma \ref{lem:zeno}, we show that CRN is the optimal CWD coupling on the $n$th iteration of Zeno's walk.
\begin{lemma}\label{lem:zeno}
	CRN is an optimal CWD coupling on the $n$th iteration of Zeno's walk when $k$ is a QAC metric. 
	\begin{proof}
		First we show that CRN is an optimal Wasserstein distance coupling on the first iteration: We can re-write $X_n=f^{-}(X_{n-1}) + B_n (f^{+}(X_{n-1})-f^{-}(X_{n-1}))$.
		Note that $f^{+}(x)-f^{-}(x)>0$ for all $x\in\mathbb{R}$ (this is clear from Figure \ref{fig:zenoswalk}).
		This means that for any $X_{n-1}\in \mathbb{R}$, the function $X_n=f^{-}(X_{n-1}) + B_n (f^{+}(X_{n-1})-f^{-}(X_{n-1}))$ is an increasing function of $B_n$ and so by Theorem \ref{prop:incr1},  for $k$ QAC, CRN is an optimal Wasserstein distance coupling on the first iteration of Zeno's walk.
		
		Next, since $f^{+}(x), f^{-}(x)$ are increasing functions of $x$ and $B_n,1-B_n\geq 0$, the function $B_n f^{+}(x)+(1-B_n)f^{-}(x)$ is an increasing function of $x$ and so, by Theorem \ref{thm:crnmc} CRN is the optimal CWD coupling for the $n$th iteration of the MC.   
	\end{proof}
\end{lemma}

\begin{figure}
	\centering
	\includegraphics[width=0.7\linewidth]{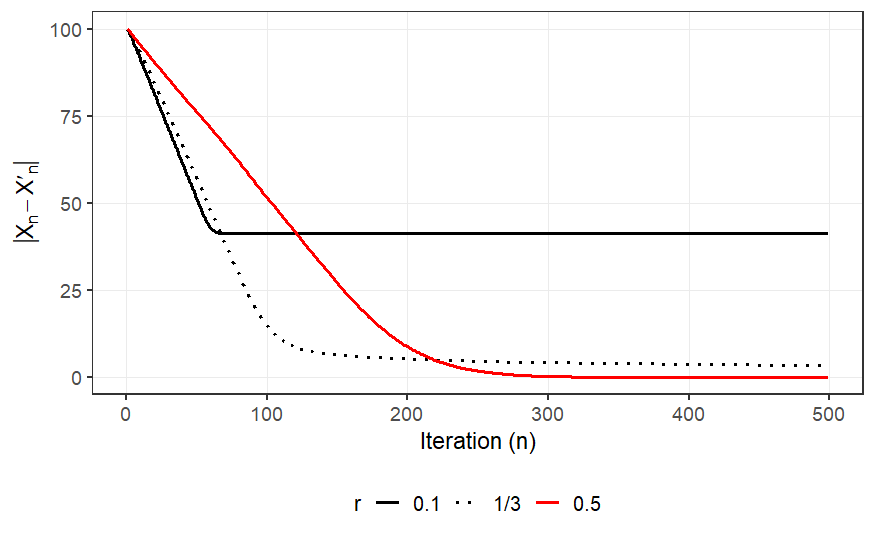}
	\caption{The simulated expected value of $|X_n-X'_n|$ where $X_n,X'_n$ follow Zeno's walk and are simulated using CRN. The CRN
		estimate uses the mean of $1000$ simulations. $X_0=0,X'_0=100$.
		}
	\label{fig:zenocrn}
\end{figure}

Figure \ref{fig:zenocrn} displays the expected distance using CRN between two copies of the Zeno walk, $X_n$ and $X'_n$, initialized at $X_0 = 0$ and $X'_0 = 100$, respectively. 

Moreover, Section 7.2 of \cite{localcont} asserts that the process fails to converge whenever $r \leq 1/3$. This theoretical result is reflected in Figure \ref{fig:zenocrn}: the black curves (corresponding to $r = 0.1$ and $r = 1/3$) are non-convergent, whereas the red curve (corresponding to $r = 0.5$) converges.

\subsection{Autoregressive normal process in $\mathbb{R}^q$}
\begin{example}[Autoregressive normal process in $\mathbb{R}^q$]\label{ex:arnormdep}
	The random vector $\{\vecx_n\}_{n\geq 1}\in \mathbb{R}^q$ is an autoregressive normal process if for matrix $A\in \mathbb{R}^{q\times q}$ and normally distributed random vector $\vec{\theta}_n\sim N(\vec{0}_q,\Sigma)$, $\Sigma$ is the covariance matrix, 
	$$\vecx_{n}=A\vecx_{n-1}+\vec{\theta}_n$$
\end{example}

If $A$ has only positive entries, then by Lemma \ref{lem:arnormdepnstep} below, we show that CRN is an optimal CWD coupling for the $n$th iteration of the MC.

\begin{lemma}\label{lem:arnormdepnstep}
	For the autoregressive normal process in $\mathbb{R}^q$, CRN is an optimal CWD coupling at the first iteration. If $A$ has non-negative entries, that is, $a_{lj}\geq0$ for all $l,j\in{1,\ldots,q}$, then CRN is an optimal CWD coupling at every iteration $n\geq1$. In particular,
	\begin{equation}\label{eq:arnormdepnstep}
		\begin{aligned}
			&\widetilde{W}_{||\cdot||_p,p}
			\bigl(\mathcal{L}(\vec{X}_n),\mathcal{L}(\vec{Y}_n)\bigr)\
			=
			\left(
			\mathbb{E}\left[
			||f_n(\Theta_n,\vec{x}_0)
			-f_n(\Theta_n,\vec{y}_0)||_p^p
			\right]
			\right)^{1/p}.
		\end{aligned}
	\end{equation}
\end{lemma}

	\begin{proof}
		
		CRN is the optimal CWD coupling at the first iteration: For any fixed $i\in\{1,\ldots,q\}$ and initial value $\vec{x}_0$, the coordinate $X_{1,i}=(A\vec{x}_{0})_{i}+\theta_{1,i}$ is an increasing function of $\theta_{1,i}$. So, by Theorem \ref{thm:rdrp}, CRN is an optimal CWD coupling for the first iteration of the autoregressive normal process in $\mathbb{R}^q$.
		
	CRN is the optimal CWD coupling at the $n$th iteration:
	Now suppose that $A$ has non-negative entries. For any $\vec x,\vec y\in\mathbb R^q$ such that $\vec x\leq\vec y$ coordinatewise, we have
$
	A\vec x+\vec\theta\leq A\vec y+\vec\theta
$
	coordinatewise, since $a_{lj}\geq0$ for all $l,j\in{1,\ldots,q}$. Hence, $f(\vec\theta,\vec x)=A\vec x+\vec\theta$ is coordinatewise nondecreasing in $\vec x$. Moreover, $f$ is coordinatewise nondecreasing in $\vec\theta$, since each component of $\vec\theta$ enters the corresponding component of $f$ additively.
	
	Thus, the assumptions of Theorem \ref{thm:crnmc} are satisfied, and CRN is an optimal CWD coupling at every iteration $n\geq1$. Equation \ref{eq:arnormdepnstep} follows.

	\end{proof}


\subsection{Bayesian location model Gibbs sampler}
\begin{example}[Bayesian location model Gibbs sampler]\label{ex:locmodel}
	Suppose that we are given data points $Y_1,\ldots, Y_J \sim N(\mu, \tau^{-1})$ where $J\geq 3$ and $\mu, \tau^{-1}$ are unknown with flat priors on $\mathbb{R}$ and $\mathbb{R}_+$, respectively. The Gibbs algorithm is based on the following conditional posterior distributions of $\mu$ and $\tin$.
	\begin{align*}
		\mu_{n+1} = \bar{y} + Z_{n+1}/\sqrt{J \tau_{n}} && \tau^{-1}_{n+1} = \frac{S+J(\bar{y}-\mu_{n+1})^2}{2G_{n+1}}
	\end{align*}
	Where $Z_{n}\sim N(0,1)$ and $G_{n}\sim \Gamma(\frac{J}{2}+1,1)$ are independent and $S=\sum_{i=1}^J (y_i-\bar{y})^2$. $\Gamma(\alpha, \beta)$ represents the gamma distribution, $\Gamma^{-1}(\alpha, \beta)$ represents the inverse gamma distribution where $\alpha$ and $\beta$ represent the shape and scale, respectively.
	
	The variable $\tau^{-1}$ can be written as an IRF: $\tinv_{n}=H_nG_n\tinv_{n-1}+G_n$
	where $H_n\sim \Gamma\left(\frac{1}{2}, \frac{S}{2}\right)$ and $G_n\sim \Gamma^{-1}\left(\frac{J+2}{2}, \frac{S}{2}\right)$ are independent.
\end{example}

The following lemma states that CRN is also an optimal CWD coupling for the $n$th iteration of the Bayesian location model Gibbs sampler.
	\begin{lemma}\label{lem:locmodel}
		CRN is an optimal CWD coupling on the $n$th iteration of the Bayesian location model Gibbs sampler when $k$ is a QAC metric. That is for i.i.d. $\vec{\theta}_n=(H_n,G_n)$,
		\begin{align}\label{eq:locmodel}
			\widetilde{W}_{k}(\mathcal{L}(\tau^{-1}_{n}), \mathcal{L}(\tau^{-1'}_{n}))=
			E[k(f_n(\Theta_n,\tau^{-1}_{0}),f_n(\Theta_n,\tau^{-1'}_{0}) )]
		\end{align}
		\begin{proof}
			Note first that $H_n, G_n,\tau_{n-1}^{-1}\geq 0$. 
			For any $H_n=h, \tau_{n-1}^{-1}=\tinv$, the function $\tinv_{n}=hG_n\tinv+G_n$ is increasing with respect to $G_n$. 
			For any $G_n=g, \tau_{n-1}^{-1}=\tinv$, the function $\tinv_{n}=H_ng\tin+g$ is increasing with respect to $H_n$. 
			For any $H_n=h, G_n=g$, the function $\tinv_{n}=hg\tin_{n-1}+g$ is increasing with respect to $\tin_{n-1}$. Thus, $f$ is coordinatewise monotone in terms of $(H_n, G_n)$ and $\tau^{-1}_n$.
			
			So by Theorem \ref{thm:crnmc}, CRN is an optimal CWD coupling on the $n$th iteration of the Bayesian location model Gibbs sampler when $k$ a QAC metric  and Equation \ref{eq:locmodel} follows.
			
		\end{proof}
	\end{lemma}


\subsection{Linear ARCH process}
\begin{example}[Linear ARCH process]\label{ex:lineararch}
	Let $\{X_n\}_{n\geq 1}\in \mathbb{R}$ be a linear ARCH process. Then for i.i.d. $\theta_n$
	$$X_n = (\beta_0 +\beta_1 X_{n-1})\theta_n$$
\end{example}

We show that for the linear ARCH process, if $X_n\geq 0$ for all $n\in\mathbb{N}$, CRN is the joint coupling that generates the CWD for iteration $n\geq 1$. 
\begin{lemma}\label{lem:lineararch}
	Suppose that $\beta_0,\beta_1, \theta_n>0$ a.s. and $x_0,y_0>0$, then CRN is an optimal CWD coupling on the $n$th iteration of the linear ARCH process when $k$ is a QAC metric. 
	\begin{align}\label{eq:lineararch}
		\widetilde{W}_{k}(\mathcal{L}(X_{n}), \mathcal{L}(Y_{n}))=
		E[k(f_n(\Theta_n, x_0),f_n(\Theta_n, y_0))]
	\end{align}
	
	\begin{proof}
		Since $(\beta_0 +\beta_1 X_{n-1})>0$ w.p. 1, $X_n=f(\theta_n,X_{n-1})$ is always an increasing function of $\theta_n$. Since $\beta_1\theta_n>0$ w.p. 1, $f(\theta_n,X_{n-1})$ is always an increasing function of $X_{n-1}$. So by Theorem \ref{thm:crnmc}, CRN is an optimal CWD coupling on the $n$th iteration of the linear ARCH process when $k$ is a QAC metric. 
	\end{proof}
\end{lemma}


\section{When monotone conditions are not satisfied}\label{sec:nonmon}

Monotonicity is important when establishing the optimal CWD coupling between two random variables, because it orders the function with regard to what is relevant: the distance between $f(\theta,x)$ and $f(\theta,y)$ as a function of $U\sim Unif(0,1)$ where $\theta=F^{-1}_{\theta}(U)$. However, sometimes $f$ is not monotone. In such cases we ask ourselves, is CRN the optimal CWD coupling? Probably not. So, how could we generate a coupling that has lower expected distance? If we can propose an equimeasurable rearrangement of $f$ w.r.t $U$, $f^*$, that is monotone, then CRN applied to the proposed $f^*$ would generate the optimal CWD coupling.

In the following section, we propose two equimeasurable transformations $g:[0,1]\to [0,1]$ and $h:[0,1]\times \{0,1\}\to [0,1]$ that, when used correctly, generate monotone functions and consequently lower expected distances when simulated using CRN. 

\subsection{Equimeasurable functions}
`Equimeasurable' when used within the context of CRN was discussed as early as 1934 \cite{inequalities}. The equimeasurable functions, however, were limited to the quantile transformation of $U\sim \unif(0,1)$, which is described in Section \ref{sec:invdf}. 
Equimeasurable rearrangements were also defined in Chapter 1 of \cite{villani}, and \cite{quantile1}, but was again only defined in terms of quantile functions. 

While the quantile function is an optimal transformation, in the problem setup with IRFs we do not know the CDF of $X$, but only know that $X=f(\theta,x)$ and that $\theta=F^{-1}_{\theta}(U)$ for known $F_{\theta}$. Thus we're interested in generating an equimeasurable monotone rearrangment of $f(F^{-1}_{\theta}(U),x)$ with respect to $U$ by creating a mapping $g:[0,1]\to [0,1]$ such that $f(F^{-1}_{\theta}(g(U)),x)$ is a monotone function of $U$ and $\mathcal{L}(f(F^{-1}_{\theta}(g(U)),x))=\mathcal{L}(f(F^{-1}_{\theta}(U),x))$. 

In this section we study equimeasurable functions as transformations on the random variable $U$ that preserve the measure of $U$ (see Definition \ref{def:equimeasurable}). This is similar to the approach taken in \cite{equimeasurable, equimeasurable2}, which studied generating lower bounds on eigenvalues using rearrangements of the eigenfunctions. 

\begin{definition}[Equimeasurable function of $f(U)$]\label{def:equimeasurable}
	Let $f:[0,1]\to \mathbb{R}$ be a measurable function and $U\sim \unif(0,1)$. $f^*$ is an equimeasurable function of $f$ if $\mathcal{L}(f(U))=\mathcal{L}(f^*(U))$.
\end{definition} 

An equimeasurable function $f^*$ is specific to the function $f$ and the random variable $U$. In this paper we restrict our definition to $U\sim \unif(0,1)$. 

The following are two function transformations that can be applied to $U$. We can then use these transformations to generate monotone equimeasurable transformations of our random function. 

If the values of a function are flipped over an interval $(a,b)\in[0,1]$, Lemma \ref{lem:intflip} states that this is an equimeasurable function.

Lemma \ref{lem:intmerge} states that if I partition an interval into two parts $(a,b)=(a,t)\cup(t,b)$ then I can replace the interval $(a,b)$ with $(a,t)$ with probability $\frac{t-a}{b-a}$ and replace it with $(t,b)$ with probability $\frac{b-t}{b-a}$, hence `merging' two intervals.

\begin{lemma}[Interval flip]\label{lem:intflip}
	Let $(a,b)\subset [0,1]$, $U\sim\unif(0,1)$, and the function $g$ to be 
	\begin{equation}\label{eq:g}
		g(U)=
		\begin{cases}
			U & \text{if } U\notin (a,b)\\
			a+b-U & \text{if } U\in (a,b)
		\end{cases}
	\end{equation}
	 Then $\mathcal{L}(f(U))=\mathcal{L}(f(g(U)))$. 
	The proof can be found in Section \ref{sec:intflip}.
\end{lemma} 

\begin{lemma}[Interval merge]\label{lem:intmerge}
	Let $(a,b)\subset [0,1]$, $t\in (a,b)$, $U\sim \unif (0,1)$ and the function $h(U,\xi)$ be as follows for independent random variables $U\sim \unif (0,1)$ and $\xi\sim \mathrm{Bernoulli}(\frac{t-a}{b-a})$
	\begin{equation}\label{eq:h}
		h(U,\xi)=
		\begin{cases}
			\xi \left(\frac{t-a}{b-a}(U-a)+a\right) + (1-\xi) \left(\frac{b-t}{b-a}(U-a)+t\right) & \text{if } U\in (a,b)\\
			U & \text{if } U\notin (a,b)
		\end{cases}
	\end{equation}
	Then $\mathcal{L}(f(U))=\mathcal{L}(f(h(U,\xi)))$.
	The proof can be found in Section \ref{sec:intmerge}.
\end{lemma}

\subsection{Example}\label{subsec:ranlogmap2}

\begin{numexmp}[Random logistic maps]\label{numex:ranlogmap}
We revisit Example \ref{ex:randomlogmap}, to prove by counterexample that CRN is not always the optimal CWD coupling for a two-step random logistic map, even though it is always the optimal CWD one step coupling. Suppose that the initial values of the random logistic map are 
$X_0=\frac{1}{2}$ and $Y_0=\frac{1}{8}$, then the second iteration of the functions are
\begin{align}\label{eq:x2y2}
	X_2&=4\beta_2\beta_1(1-\beta_1) &
	Y_2&=\frac{7}{4}\beta'_2\beta'_1\left(1-\frac{7}{16}\beta'_1\right)
\end{align} 

Recall $\beta_1,\beta_2\sim \text{Beta}(\alpha+1/2,\alpha-1/2)$. Denote $F_{\beta}$ to be the CDF of $\text{Beta}(\alpha+1/2,\alpha-1/2)$ so that $F_{\beta}^{-1}(U)\sim \text{Beta}(\alpha+1/2,\alpha-1/2)$. We can then re-write $X_2$ as follows where $U_1\sim \unif(0,1)$ and $U_1, \beta_2$ are independent,
\begin{align*}
	X_2(U_1,\beta_2)&=4\beta_2F^{-1}_{\beta}(U_1)(1-F^{-1}_{\beta}(U_1))
\end{align*} 
Observe that for any $\beta_2$, $X_2$ is an increasing function of $U_1$ when $U_1\in [0,F_{\beta}(0.5)]$ and a decreasing function of $U_1$ when $U_1\in [F_{\beta}(0.5),1]$. We will construct an equimeasurable transform of $X_2(U_1, \beta_2)$, $X''_2(U_1, \beta_2, \xi)$, that is monotone in $U_1$ for $\xi\in \{0,1\}$.  To do so, we first generate an equimeasurable function, $X'_2(U_1,\beta_2)=X(g(U_1),\beta_2)$, that is increasing over $U_1\in (F_{\beta}(0.5),1]$ where $g:[0,1]\to[0,1]$ is defined as follows,
 
\begin{equation}\label{eq:g}
	g(U)=
	\begin{cases}
		U & \text{if } U\in [0,F_{\beta}(0.5)]\\
		1+F_{\beta}(0.5)-U & \text{if }U\in (F_{\beta}(0.5),1]\\
	\end{cases}
\end{equation}
By Lemma \ref{lem:intflip}, $\mathcal{L}(X'_2)=\mathcal{L}(X_2)$. 

The interval flip is easiest to visualize in Figure \ref{fig:ranlogmapflip}.  In graph (A), when $U_1<F_{\beta}(0.5)\approx 0.288$ ($\beta_1<0.5$), $X_2(U_1,\beta_2)=X'_2(U_1,\beta_2)$, however when $U_1>F_{\beta}(0.5)$ ($\beta_1>0.5$), $X_2(U_1,\beta_2)$ is represented by the black line while the transform $X'_2(U_1, \beta_2)$ is represented by the red line. Note that the transform  $X'_2(U_1, \beta_2)$ is increasing when $U_1>F_{\beta}(0.5)$.

\begin{figure}
	\centering
	\includegraphics[width=1\linewidth]{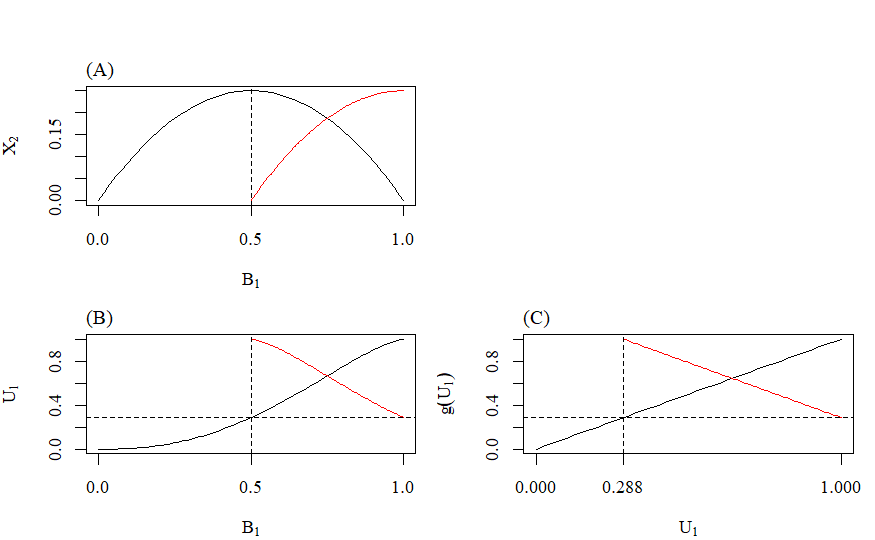}
	\caption{Set $\alpha=2$ so that $\beta_1\sim \text{Beta}(2.5,1.5)$ and $F_{\beta}(0.5)= 0.288$. Set $\beta_2=1$ (its value is of little importance since it does not affect whether $\beta_1$ is increasing or not). In graph (A), the black line denotes $X_2(U_1,1)$ and the red line denotes $X'_2(U_1,1)$ when it is not equal to $X_2(U_1,1)$. 
	Graph (C) represents the transformation $g(U_1)$ in red compared to the identity function. Graph (B) then imposes the flip that was applied by $g$ onto the values of $\beta_1$. Graph (A) shows the effect of the flip, $g$, through $\beta_1$ onto the value of $X_2$.}
	\label{fig:ranlogmapflip}
\end{figure}

The transform $X'_2(U_1, \beta_2)$ is increasing over $U_1\in [0,F_{\beta}(0.5))$ and $U_1\in (F_{\beta}(0.5),1]$. However, it is not monotonically increasing. To generate an equimeasurable map that is monotonically increasing with respect to $U$, we use Lemma \ref{lem:intmerge} and introduce the auxiliary random variable $\xi$. 

Let $(a,b)=[0,1]$ and $t=F_{\beta}(0.5)$. Then $X''_2(U_1,\beta_2,\xi)=X_2(g(h(U_1,\xi)),\beta_2)$ where $h(U_1,\xi)$ is generated according to Equation \ref{eq:h} is an equimeasurable function of $U$ (Lemma \ref{lem:intmerge}). To summarize, $X''_2$ can be defined as follows for $\xi\sim \text{Bernoulli}(F_{\beta}(0.5))$
\begin{equation}\label{eq:X2gh}
	X''_2(U_1,\beta_2,\xi)=
	\begin{cases}
		4\beta_2 F_{\beta}^{-1}(1-U_1(1-F_{\beta}(0.5)))[1-F_{\beta}^{-1}(1-U_1(1-F_{\beta}(0.5)))] & \text{if } \xi=0\\
		4\beta_2 F_{\beta}^{-1}(F_{\beta}(0.5) U_1) [1-F_{\beta}^{-1}(F_{\beta}(0.5) U_1)] & \text{if } \xi=1\\
	\end{cases}
\end{equation}
  Lemmas \ref{lem:intmerge} and \ref{lem:intflip} imply that  $X''_2(U_1, \beta_2,\xi)$ is an equimeasurable transform of $X_{2}(U_1, \beta_2)$. Figure \ref{fig:ranlogmapdensities} (A) graphs the simulated density of $X_2(U_1, \beta_2)$, $X'_2(U_1, \beta_2)$ and $X''_2(U_1, \beta_2,\xi)$, which visually shows that the  densities are equal. 

\begin{figure}
	\centering
	\includegraphics[width=0.7\linewidth]{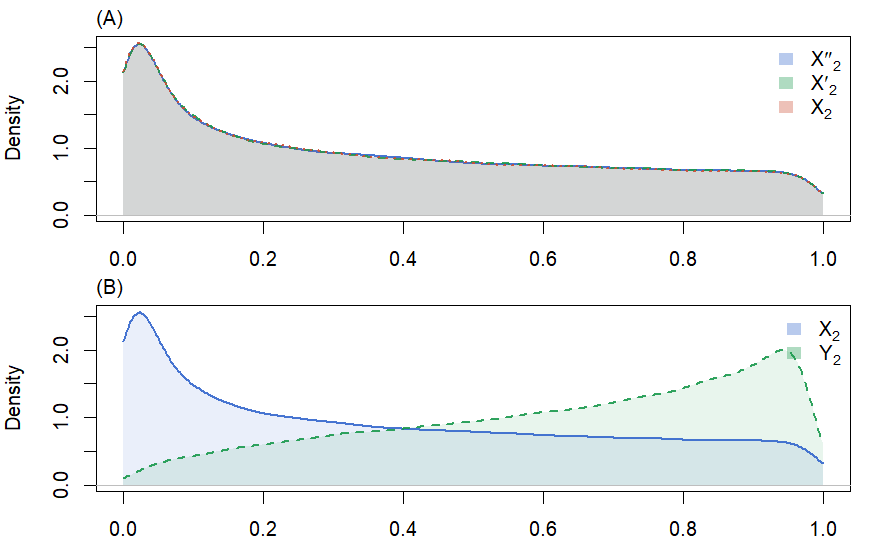}
	\caption{(A): Simulated density of $X_2=4\beta_2\beta_1(1-\beta_1)$, $X'_2(U_1,\beta_2)$, and $X''_2(U_1,\beta_2,\xi)$ (see Equation \ref{eq:X2gh}). The random variables $\beta_i\sim \text{Beta}(1.5,0.5), i\in \{1,2\}$, $U_1\sim \unif(0,1)$, $\xi\sim \text{Bernoulli}(F_{\beta}(0.5))$ are independent. 
This shows that the random variables $X'_2$, which is based on the transformation $g$ and $X''_2$, which is based on the transformations $g$ and $h$ are indeed equimeasurable to $X_2$. (B): Comparison of the density of $X_2$ and $Y_2$
	}
	\label{fig:ranlogmapdensities}
\end{figure}

Figure \ref{fig:randomlogmapcrnvsmin} shows how the solid line, which is the expected value of $|X_2(U_1,\beta_2)-Y_2(U_1,\beta_2)|$ simulated using CRN has a larger expected distance compared to the transformed expected value of $|X'_2(U_1,\beta_2,\xi)-Y_2(F^{-1}_{\beta}(U_1),\beta_2)|$ simulated using CRN. This shows that $E[|X_2(U_1, \beta_2)-Y_2(U_1, \beta_2)|]$ is strictly greater than the CWD.
\begin{align*}
	\widetilde{W}_{|\cdot|}(\mathcal{L}(X), \mathcal{L}(Y))
	\leq E[|X'_2(U_1, \beta_2)-Y_2(U_1, \beta_2)|]
	<
	E[|X_2(U_1, \beta_2)-Y_2(U_1, \beta_2)|]
\end{align*}

Figure \ref{fig:randomlogmapcrnvsmin} shows how the long-dashed line, which is the expected value of $|X'_2(U_1,\beta_2)-Y_2(U_1,\beta_2)|$ simulated using CRN has a larger expected distance compared to the dotted line, the transformed expected value of $|X''_2(U_1,\beta_2,\xi)-Y_2(U_1,\beta_2,\xi)|$ simulated using CRN. This shows that $E[|X'_2(U_1, \beta_2)-Y_2(U_1, \beta_2)|]$ is strictly greater than the Wasserstein distance.
\begin{align*}
	W_{|\cdot|}(\mathcal{L}(X), \mathcal{L}(Y))
	\leq E[|X''_2(U_1, \beta_2,\xi)-Y_2(U_1, \beta_2,\xi)|]
	<
	E[|X'_2(U_1, \beta_2)-Y_2(U_1, \beta_2)|]
\end{align*}

Note that we cannot use $E[|X''_2(U_1, \beta_2,\xi)-Y_2(U_1, \beta_2,\xi)|]$ to bound the CWD from above, because the joint distribution that is generated by adding $\xi$ may not be contained in the set of coordinate-independent joint distributions generated $\mathcal{J}_{IND}(\mu,\mu)$ where $\mu=\mathcal{L}(U_1,\beta_2)$.

\begin{figure}
	\centering
	\includegraphics[width=0.7\linewidth]{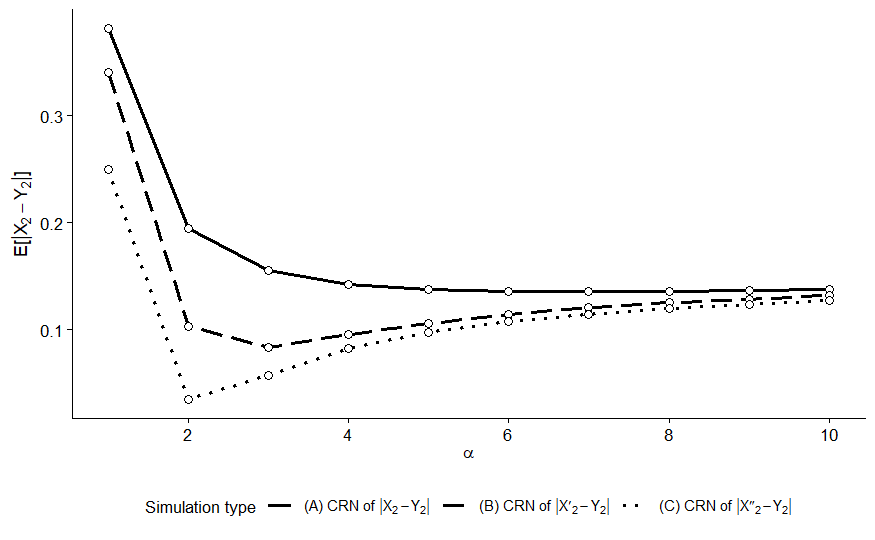}
	\caption{Line (A) represents the simulated expected distance of $|X_2(U_1,B_2)-Y_2(U_1,B_2)|$ ie. the expected distance when CRN is applied to the original function. Line (B) represents the simulated expected distance of $|X'_2(U_1,B_2)-Y_2(U_1,B_2)|$ ie. the expected distance when CRN is applied to the transformed $X'_2(U_1,B_2,\xi)$. Line (C) represents the simulated expected distance of $|X''_2(U_1,B_2,\xi)-Y_2(U_1,B_2)|$ ie. the expected distance when CRN is applied to the transformed $X''_2(U_1,B_2,\xi)$.}
	\label{fig:randomlogmapcrnvsmin}
\end{figure}

\end{numexmp}

	\section{Appendix}
	\addtocontents{toc}{\protect\setcounter{tocdepth}{0}}
	
	\subsection{Proof of Proposition \ref{thm:equivinv4}}\label{sec:skorohod}

\begin{proof}{Proof of Proposition \ref{thm:equivinv4}}\label{pf:skorohod}
	
	$\impliedby$: $|F^{-1}_{n}(U)-F^{-1}(U)| \to 0 \text{ w.p. 1}$ means that $F^{-1}_{n}(U) \to F^{-1}(U)$ almost surely, which means that $\mathcal{L}(F^{-1}_{n}(U)) \to \mathcal{L}(F^{-1}(U))$ in distribution. Since $F^{-1}_{n}(U)\sim \mathcal{L}(X_n)$ and $F^{-1}(U)\sim\mathcal{L}(X_{\infty})$, this implies 
	$\mathcal{L}(X_n)\Rightarrow \mathcal{L}(X_{\infty})$.
	
	$\implies$
	This proof greatly resembles the proof of Theorem 10.1.1 (4) of \cite{firstlook}.
	
	Denote $D_{F^{-1}}\in [0,1]$ as the set of discontinuous points on $F^{-1}$ and $C_{F^{-1}}\in [0,1]$ as the set of continuous points on $F^{-1}$.
	
	Let $0<u<1$,  $\epsilon>0$, and choose $\omega$ such that $F^{-1}(u)-\epsilon<\omega<F^{-1}(u)$ and $\pi(\omega)=0$. Then $F(\omega)<u$ and since $F_{n}(\omega)\to F(\omega)$, for sufficiently large $n$, $F_{n}(\omega)<u$ and so, $F^{-1}(u)-\epsilon<\omega<F^{-1}_{n}(u)$. Thus,
	\begin{equation}\label{eq:skorohodinf}
		\lim \inf F_{n}^{-1}(u)\geq F^{-1}(u)
	\end{equation}
	
	Let $0<u<u'<1$ and choose $\omega$ such that $F^{-1}(u')<\omega<F^{-1}(u')+\epsilon$ and $\pi(\omega)=0$. Then $u<u'\leq F(F^{-1}(u'))\leq F(\omega)$ and since $F_{n}(\omega)\to F(\omega)$, for sufficiently large $n$, $u\leq F_{n}(\omega)$ and $F^{-1}_{n}(u)\leq \omega < F^{-1}(u')+\epsilon$. Thus,
	\begin{equation}\label{eq:skorohodsup}
		\lim \sup F_n^{-1}(u)\leq F^{-1}(u')
	\end{equation}
	
	By equation \ref{eq:skorohodinf} and equation \ref{eq:skorohodsup} if $F^{-1}$ is continuous at $u$ (ie. $u\in C_{F^{-1}}$), then $\lim_{n\to \infty} F^{-1}_{n}(u)= F^{-1}(u)$. Or in other words, $P(\lim_{n\to \infty} |F^{-1}_{n}(u)- F^{-1}(u)|=0\mid U\in C_{F^{-1}})=1$. 
	
	Next suppose that $U\sim Unif(0,1)$. Since $F^{-1}$ is non-decreasing on $(0,1)$, it has countably many discontinuous points and so $P(U\in D_{F^{-1}})=0$ (and $P(U\in C_{F^{-1}})=1$).
	
	Thus, the probability of $\lim_{n\to \infty} |F^{-1}_{n}(u)- F^{-1}(u)|=0$ is as follows,
	\begin{align*}
		&P(\lim_{n\to \infty} |F^{-1}_{n}(U)- F^{-1}(U)|=0)\\
		&\geq P( \lim_{n\to \infty} |F^{-1}_{n}(U)- F^{-1}(U)|=0\mid U\in C_{F^{-1}})P(U\in C_{F^{-1}}) \\
		&=1
	\end{align*}
	
	And so, when CRN is applied to the quantile function, $|F^{-1}_n(U)-F^{-1}(U)|\to 0$ w.p. 1.
\end{proof}
	
	\subsection{Proof of Theorem \ref{thm:wasseqinf}}\label{pf:wasseqinf}
	
	For simplicity, we remove the vector notation from the random variables (ex. $\vec{X}$ becomes $X$). 
	Define  $X=f(\theta,x_0)=g(\theta)$ and $Y=f(\theta',y_0)=h(\theta')$. Recall that $\theta,\theta'\sim \mu$. Denote $\nu\in \mathcal{J}(\mathcal{L}(X),\mathcal{L}(Y))$. Denote $\mathcal{X}^d = \prod_{i=1}^d\mathcal{X}_i$.
	
	By Proposition 10.4.12 of \cite{Bog07}, since $\theta,\theta'\sim \mu$ and $g,h$ are Borel mappings, then there exists Borel regular conditional probability measures $\mu(\cdot\mid x)=\mu(\cdot\mid g(\theta)=x), x\in X$ on $\theta$ and $\mu(\cdot\mid y)=\mu(\cdot\mid h(\theta')=y), y\in Y$ on $\theta'$.
	
	\begin{definition}\label{def:gamma}
		Define $\gamma:\mathcal{B}(\mathcal{X}^{d}\times \mathcal{X}^d)\to \mathbb{R}$ such that for any $C\in \mathcal{B}(\mathcal{X}^{d}\times \mathcal{X}^{d})$, 
		\begin{align*}
			\gamma(C)
			&=\int_{\mathbb{R}^q\times \mathbb{R}^q}\int_{C} \mu(\d\theta\mid x) \mu(\d \theta'\mid y)\nu(\d x, \d y)
		\end{align*}
		
		Note that if $C=A\times B$ where $A,B\in \mathcal{B}(\mathcal{X}^{d})$
		\begin{align*}
			\gamma(C)
			&
			=\int_{\mathbb{R}^q\times \mathbb{R}^q} \mu(A\mid x) \mu(B\mid y)\nu(\d x, \d y)
		\end{align*}
	\end{definition}
	
	\begin{lemma}\label{lem:gammaprobmeasure}
		$\gamma$ is a probability measure on $\mathcal{B}(\mathcal{X}^{d}\times \mathcal{X}^{d})$.
		\begin{proof}
		\begin{enumerate}
			\item $\gamma(\mathcal{X}^{d}\times \mathcal{X}^{d})=1$
			\begin{align*}
				\gamma(\mathcal{X}^{d}\times \mathcal{X}^{d})
				&= \int_{\mathbb{R}^q\times \mathbb{R}^q} \mu(\mathcal{X}^{d}\mid x) \mu(\mathcal{X}^{d}\mid y)\nu(\d x, \d y)\\
				&= \int_{\mathbb{R}^q\times \mathbb{R}^q} \nu(\d x, \d y)&\text{Since $\mu(\mathcal{X}^{d}\mid x)=\mu(\mathcal{X}^{d}\mid y)=1$}\\
				&=1
			\end{align*}
			\item $\gamma(\emptyset)=0$
			\begin{align*}
				\gamma(\emptyset)= \int_{\mathbb{R}^q\times \mathbb{R}^q}\mu(\emptyset\mid x) \mu(\emptyset \mid y)\nu(\d x, \d y)= 0 \int_{\mathbb{R}^q\times \mathbb{R}^q} \nu(\d x, \d y)=0\times 1=0
			\end{align*} 
			\item Suppose that $A_{1}, A_{2}, \ldots \in \mathcal{B}(\mathcal{X}^{d}\times \mathcal{X}^{d})$ is a sequence of disjoint sets, then  $\gamma(\dot{\cup}_{i}^{\infty}A_{i})=\sum_{i=1}^{\infty}\gamma(A_{i})$
			\begin{align*}
				\gamma(\dot{\cup}_{i}^{\infty}A_i)&= \int_{\mathbb{R}^q\times \mathbb{R}^q} \int_{\dot{\cup}_{i=1}^{\infty}A_{i}} \mu(\d\theta\mid x) \mu(\d\theta'\mid y)\nu(\d x, \d y)\\
				&= \sum_{i=1}^{\infty} \int_{\mathbb{R}^{q}\times \mathbb{R}^{q}} \int_{A_{i}} \mu(\d\theta\mid x) \mu(\d\theta'\mid y)\nu(\d x, \d y) &\text{by MCT (Exercise 1.5.6 of \cite{durrett})}\\
				&= \sum_{i=1}^{\infty}  \gamma(A_{i})
			\end{align*}
		\end{enumerate}
		Thus by definition of a probability measure (Section 2.1 of \cite{firstlook}), $\gamma$ is a probability measure.
	\end{proof}
	\end{lemma}
	
	\begin{lemma}\label{lem:gammamarginals}
	The marginals of $\gamma$ are $\mu$. That is, $\gamma(A\times \mathcal{X}^{d})=\mu(A)$ for $A\in\mathcal{B}(\mathcal{X}^{d})$.
	\begin{proof}
		\begin{align*}
			\gamma(A\times \mathcal{X}^{d})&=\int_{\mathbb{R}^q\times \mathbb{R}^q}\int_{\mathcal{X}^{d}}\int_{A}\mu(d\theta\mid x) \mu(d\theta'\mid y) \nu(\d x,\d y)\\
			&=\int_{\mathbb{R}^q\times \mathbb{R}^q} \mu(A\mid x) \mu(\mathcal{X}^{d} \mid y) \nu(\d x,\d y)\\
			&=\int_{\mathbb{R}^q\times \mathbb{R}^q} \mu(A\mid x) \nu(\d x,\d y)\\
			&=\mu(A)
		\end{align*}
		
		By similar reasoning, $\gamma(\mathcal{X}^{d}\times A)=\mu(A)$.
	\end{proof}
	\end{lemma}
	
	\begin{lemma}\label{lem:gammapushforward}
		 $\nu$ is the pushforward of $\gamma$. That is, $(g\times h)_{*}\gamma=\nu$.
		\begin{proof}
		By definition, for any $B\in \mathcal{B}(\mathbb{R}^q\times \mathbb{R}^q)$ $(g\times h)_{*}\gamma(B)=\gamma((g\times h)^{-1}(B))$ (Definition 2.40 of \cite{bridgeford2025measure}). So we must show that for any  $B\in \mathcal{B}(\mathbb{R}^q\times \mathbb{R}^q)$, $\gamma((g\times h)^{-1}(B))=\nu(B)$. Let $B\in \mathcal{B}(\mathbb{R}^q\times \mathbb{R}^q)$,
		\begin{align}\label{eq:pushforgamma}
			(g\times h)_{*}\gamma(B)&=\gamma((g\times h)^{-1}(B))\\
			&=\int_{\mathcal{X}^{d}\times \mathcal{X}^{d}}\int_{\mathbb{R}^q\times \mathbb{R}^q} I[(\theta,\theta')\in (g\times h)^{-1}(B)] \mu(\d\theta\mid x) \mu(\d\theta'\mid y)\nu(\d x, \d y)
		\end{align}
		For fixed $x,y$ note that for all $\theta\sim \mu(\cdot \mid x) $ and $\theta'\sim \mu(\cdot \mid y)$, either $I[(\theta,\theta')\in (g\times h)^{-1}(B)]=1$ or $I[(\theta,\theta')\in (g\times h)^{-1}(B)]=0$ depending on whether $(g(\theta),h(\theta'))=(x,y)\in B$,  so that $I[(\theta,\theta')\in (g\times h)^{-1}(B)] =I[(x,y)\in B]$.
		
		And so Equation \ref{eq:pushforgamma} equals
		\begin{align*}
			&=\int_{\mathbb{R}^q\times \mathbb{R}^q} \int_{\mathcal{X}^{d}\times \mathcal{X}^{d}} I[(x,y)\in B] \mu(\d\theta\mid x) \mu(\d\theta'\mid y)\nu(\d x, \d y) \\
			&= \int_{\mathbb{R}^q\times \mathbb{R}^q} I[(x,y)\in B] \int_{\mathcal{X}^{d}\times \mathcal{X}^{d}} \mu(\d\theta\mid x) \mu(\d\theta'\mid y)\nu(\d x, \d y) \\
			&= \int_{\mathbb{R}^q\times \mathbb{R}^q} I[(x,y)\in B] \mu(\mathcal{X}^{d}\mid x) \mu(\mathcal{X}^{d}\mid y)\nu(\d x, \d y) \\
			&= \int_{\mathbb{R}^q\times \mathbb{R}^q} I[(x,y)\in B] \nu(\d x, \d y) \\
			&=\nu(B)
		\end{align*}
		
		Thus we have that $(g\times h)_{*}\gamma(B)=\nu(B)$ for arbitrary $B$ and so,  $(g\times h)_{*}\gamma=\nu$.
	\end{proof}
	\end{lemma}
	
	\begin{lemma}\label{lem:gammaexp}
		$E_{(\theta,\theta')\sim \gamma}[k(g(\theta),h(\theta'))]=E_{(X,Y)\sim \nu}[k(X,Y)]$
		\begin{proof}
		\begin{align*}
			&E_{(\theta,\theta')\sim \gamma}[k(g(\theta),h(\theta'))]\\
			&=\int_{\mathcal{X}^{d} \times \mathcal{X}^{d} } k(g(\theta),h(\theta'))\gamma(d\theta,d\theta')\\
			&=\int_{\mathbb{R}^q\times\mathbb{R}^q}k(x,y)((g\times h)_*\gamma)(dx,dy)&\text{By Lemma 3.12 of \cite{bridgeford2025measure}}\\
			&= \int_{\mathbb{R}^q\times \mathbb{R}^q} k[(x,y)] \nu(\d x, \d y) \\
			&= E_{(X,Y)\sim \nu}[k(X,Y)]
		\end{align*}
	\end{proof}
	\end{lemma}
	
	\begin{proof}[Proof of Theorem \ref{thm:wasseqinf}]
		Recall that $X=f(\theta,x)=g(\theta)$ and $Y=f(\theta',y)=h(\theta')$. 
		
		Note that by discussion below Proposition 2.2 of \cite{crn},
		\begin{align*}
			\mathcal{J}(\mathcal{L}(X),\mathcal{L}(Y))\supseteq \{\mathcal{L}(g(\theta),h(\theta'))\mid \theta,\theta'\sim \mu\}
		\end{align*}
		and so by Equation \ref{eq:glassyao},
		\begin{align}\label{eq:gammaupperbound}
			W_k(\mathcal{L}(X), \mathcal{L}(Y))\leq  \inf_{\gamma\in \mathcal{J}(\mu,\mu)} E_{(\theta,\theta')\sim \gamma}[k(g(\theta),h(\theta'))]
		\end{align}
		
		We want to show that the reverse is also true. Ie.
		$$W_k(\mathcal{L}(X), \mathcal{L}(Y))\geq  \inf_{\gamma\in \mathcal{J}(\mu,\mu)} E_{(\theta,\theta')\sim \gamma}[k(g(\theta),h(\theta'))]$$
		
		Let $\nu\in \mathcal{J}(\mathcal{L}(X),\mathcal{L}(Y))$ and propose $\gamma(\theta,\theta')$ as defined in Definition \ref{def:gamma}. 
		
		$\gamma\in \mathcal{J}(\mu, \mu)$ since by Lemma \ref{lem:gammaprobmeasure}, $\gamma$ is a probability measure on $\mathcal{B}(\mathcal{X}^{d}\times \mathcal{X}^{d})$ and by Lemma \ref{lem:gammamarginals}, the marginals of $\gamma$ are $\mu$. Further, by Lemma \ref{lem:gammapushforward}, $(g\times h)_{*}\gamma=\nu$ and so, for every $\nu \in \mathcal{J}(\mathcal{L}(X),\mathcal{L}(Y))$, there exists a $\gamma\in \mathcal{J}(\mu,\mu)$ such that $\mathcal{L}(g(\theta),h(\theta'))\in  \mathcal{J}(\mathcal{L}(X),\mathcal{L}(Y))$. This means that 
		\begin{align*}
			\mathcal{J}(\mathcal{L}(X),\mathcal{L}(Y))\subseteq \{\mathcal{L}(g(\theta),h(\theta'))\mid \theta,\theta'\sim \mu\}
		\end{align*}
		
		Furthermore for every $\nu\in \mathcal{J}(\mathcal{L}(X),\mathcal{L}(Y))$, the corresponding $\gamma\in \mathcal{J}(\mu,\mu)$ generates the same expected distance between $X$ and $Y$. That is, by Lemma \ref{lem:gammaexp},
		$$E_{(\theta,\theta')\sim \gamma}[k(g(\theta),h(\theta'))]=E_{(X,Y)\sim \nu}[k(X,Y)]$$
		
		And so we get the following inequality,
		\begin{align*}
			\inf_{\nu\in \mathcal{J}(\mathcal{L}(X),\mathcal{L}(Y))} E_{(X,Y)\sim \nu}[k(X,Y)]
			\geq \inf_{\gamma\in \mathcal{J}(\mu,\mu)} E_{(\theta,\theta')\sim \gamma}[k(g(\theta),h(\theta'))]
		\end{align*}
		
		Thus the Wasserstein distance is bounded below by the infimum expected distance with respect to the joint probability between $\theta,\theta'$
		\begin{align}\label{eq:gammalowerbnd}
			W_k(\mathcal{L}(X), \mathcal{L}(Y))&= \inf_{\nu\in \mathcal{J}(\mathcal{L}(X),\mathcal{L}(Y))} E_{(X,Y)\sim \nu}[k(X,Y)]\\
			&\geq \inf_{\gamma\in \mathcal{J}(\mu,\mu)} E_{(\theta,\theta')\sim \gamma}[k(g(\theta),h(\theta'))]
		\end{align}
		
		Combining equations \ref{eq:gammaupperbound} and \ref{eq:gammalowerbnd}, we get equality.
		\begin{align}
			W_k(\mathcal{L}(X), \mathcal{L}(Y))=\inf_{\gamma\in \mathcal{J}(\mu,\mu)} E_{(\theta,\theta')\sim \gamma}[k(g(\theta),h(\theta'))]
		\end{align}
		
	\end{proof}

	\subsection{Proof of Theorem \ref{prop:incr1}} \label{pf:incr1}
	\begin{proof}[Proof of Theorem \ref{prop:incr1}]
		Denote $g(\theta_1)=f(\theta_1,x)$ and $h(\theta_2)=f(\theta_2,y)$ for fixed $x$ and $y$.
		
		Since $g$ and $h$ are monotone in $\theta$ by assumption, if $\theta_1\leq \theta_1'$ and $\theta_2\leq \theta_2'$, then $g(\theta_1)\leq g(\theta_1')$ and $h(\theta_2)\leq h(\theta_2')$. By the quasi-antitone property of $k$,
		$$k\left(g(\theta_1),h(\theta_2)\right)+k\left(g(\theta_1'),h(\theta_2')\right)-k\left(g(\theta_1'),h(\theta_2)\right)-k\left(g(\theta_1),h(\theta_2')\right)\leq 0$$
		
		And so, $k(g(\theta_1),h(\theta_2))$ is quasi-antitone with respect to $\theta_1$ and $\theta_2$.
		
		Further, since $g$ and $h$ are right-continuous functions, $k(g(\theta_1),h(\theta_2))$ is right-continuous with respect to $\theta_1,\theta_2$ and so $k(g(\theta_1),h(\theta_2))$ is QAC with respect to $\theta_1,\theta_2$. We can then apply Proposition \ref{prop:interval}. 
		\begin{align*}
			\inf_{\mathcal{J}(\mu,\mu)}E[k(g(\theta_1),h(\theta_2))] 
			&=E[k(g(F^{-1}_{\theta}(U)),h(F^{-1}_{\theta}(U)))] &\text{by Prop \ref{prop:interval}}\\
			&=E[k(g(\theta),h(\theta))]\\
			&=E[k(f(\theta,x),f(\theta,y))]
		\end{align*}
		
		Combining the above equation with Theorem \ref{thm:wasseqinf}, $W_k(\mathcal{L}(X), \mathcal{L}(Y))=E[k(f(\theta,x),f(\theta,y))]$.
	\end{proof}

	\subsection{Proof of Theorem \ref{prop:rd}}\label{pf:rdinduction}
\begin{proof}
	\underline{CRN is the coordinatewise optimal CWD coupling}
	First note that by assumption, for all $i\in\{1,\ldots, d\}$ the function 
	$f(\vec{\theta},x)$ with fixed $\vec{\theta}_{-i}$ $\mu_{-i}$- a.e. is an increasing  continuous function with respect to $\theta_i$. Next since $k$ is QAC, then by Theorem \ref{prop:incr1}, for almost every pair $(\vec{\theta}_{-i},\vec{\theta}'_{-i})$ and fixed $x,y$, CRN is an optimal coupling with respect to the $i$th coordinate,
	\begin{align}\label{eq:rdcond}
		&\inf_{\gamma_i\in \mathcal{J}(\mu_i,\mu_i)} E_{(\theta_i,\theta'_i)\sim \gamma_i}[k(f((\theta_i, \vec{\theta}_{-i}),x),f( (\theta'_i, \vec{\theta}'_{-i}),y))]\\
		&=E_{\theta_i\sim \mu_i}[k(f((\theta_i, \vec{\theta}_{-i}),x),f( (\theta_i, \vec{\theta}'_{-i}),y))]
	\end{align}

	We apply a proof by induction to show that CRN is an optimal CWD coupling with respect to the unconditional expected value.
	
	\underline{Base step:} 
	
The expected value of $k(f( \vec{\theta},x), f(\vec{\theta}',y))$ conditioned on coordinates $(\theta_{2:d},\theta'_{2:d})$ is bounded below by the expected value 
	when CRN is applied to coordinate 1 ($\theta_1=\theta'_1$). This is true by equation \ref{eq:rdcond} for arbitrary joint distribution $\gamma_1\in \mathcal{J}(\mu_1,\mu_1)$ and almost every pair $(\theta_{2:d},\theta'_{2:d})$.
	\begin{align*}
		E_{(\theta_1,\theta'_{1})\sim\gamma_1}[k(f( (\theta_1, \vec{\theta}_{2:d}),x),f((\theta'_1, \vec{\theta}'_{2:d}),y))]&\geq E_{\theta_1\sim \mu_1}[k(f( (\theta_1, \vec{\theta}_{2:d}),x),f((\theta_1,\vec{\theta}'_{2:d}),y))]
	\end{align*}
	
	\underline{Induction step:} Suppose that for $0\leq N\leq d-1$, the joint expected distance conditioned on fixed coordinates $(\vec{\theta}_{N+1:d}, \vec{\theta}'_{N+1:d})$ is bounded below by the expected distance when CRN is applied to coordinates $1$ to $N$ ($\vec{\theta}_{1:N}=\vec{\theta}'_{1:N}$). Denote $\gamma_n\in \mathcal{J}(\mu_n,\mu_n)$.
	\begin{align}\label{eqn:rdbase}
		&E_{(\vec{\theta}_{1:N},\vec{\theta}'_{1:N})\sim\bigotimes_{n=1}^N \gamma_n}[k(f( (\vec{\theta}_{1:N}, \vec{\theta}_{N+1:d}),x),f((\vec{\theta}'_{1:N}, \vec{\theta}'_{N+1:d}),y))]\\
		&\hspace{1cm}\geq E_{\vec{\theta}_{1:N}\sim \bigotimes_{n=1}^N \mu_n}[k(f( (\vec{\theta}_{1:N}, \vec{\theta}_{N+1:d}),x),f((\vec{\theta}_{1:N}, \vec{\theta}'_{N+1:d}),y))]
	\end{align}
	We will then show that for $N+1$, the expected value of $k$ conditioned on fixed coordinates $(\vec{\theta}_{N+2:d}, \vec{\theta}'_{N+2:d})$ is also attained when CRN is applied to coordinates $1$ to $N+1$. 
	\begin{align*}
		&E_{(\vec{\theta}_{1:N+1},\vec{\theta}'_{1:N+1})\sim \bigotimes_{n=1}^{N+1} \gamma_n}[k(f( (\vec{\theta}_{1:N+1}, \vec{\theta}_{N+2:d}),x),f((\vec{\theta}'_{1:N+1}, \vec{\theta}'_{N+2:d}),y))]\\
		&= E_{(\vec{\theta}_{1:N},\vec{\theta}'_{1:N})\sim\bigotimes_{n=1}^N \gamma_n}[E_{(\theta_{N+1},\theta'_{N+1})\sim \gamma_{N+1}}[k(f( (\vec{\theta}_{1:N},\theta_{N+1}, \vec{\theta}_{N+2:d}),x),f((\vec{\theta}'_{1:N},\theta'_{N+1}, \vec{\theta}'_{N+2:d}),y))]]\\
		&\geq E_{(\vec{\theta}_{1:N},\vec{\theta}'_{1:N})\sim\bigotimes_{n=1}^N \gamma_n}[E_{\theta_{N+1}\sim \mu_{N+1}}[k(f( (\vec{\theta}_{1:N},\theta_{N+1}, \vec{\theta}_{N+2:d}),x),f((\vec{\theta}'_{1:N},\theta_{N+1}, \vec{\theta}'_{N+2:d}),y))]]&\text{by eq. \ref{eq:rdcond}}\\
		&= E_{\theta_{N+1}\sim \mu_{N+1}}[ E_{(\vec{\theta}_{1:N},\vec{\theta}'_{1:N})\sim\bigotimes_{n=1}^N \gamma_n}[k(f( (\vec{\theta}_{1:N},\theta_{N+1} \vec{\theta}_{N+2:d}),x),f((\vec{\theta}'_{1:N},\theta_{N+1} \vec{\theta}'_{N+2:d}),y))]]\\
		&\geq E_{\theta_{N+1}\sim \mu_{N+1}}[ E_{\vec{\theta}_{1:N}\sim\bigotimes_{n=1}^N \mu_n}[k(f( (\vec{\theta}_{1:N},\theta_{N+1} \vec{\theta}_{N+2:d}),x),f((\vec{\theta}_{1:N},\theta_{N+1} \vec{\theta}'_{N+2:d}),y))]] &\text{by eq. \ref{eqn:rdbase}}\\
		&=  E_{\vec{\theta}_{1:N+1}\sim\bigotimes_{n=1}^{N+1} \mu_n}[k(f( (\vec{\theta}_{1:N},\theta_{N+1} \vec{\theta}_{N+2:d}),x),f((\vec{\theta}_{1:N},\theta_{N+1} \vec{\theta}'_{N+2:d}),y))]
	\end{align*}
	
	The above implies that the expected value of $k$ for joint distribution $\gamma \in \mathcal{J}_{IND}(\mu,\mu) $ is bounded below by the expected value of $k$ when CRN is applied.
	\begin{align}
		E_{(\vec{\theta},\vec{\theta}')\sim \gamma}[k(f( \vec{\theta},x),f(\vec{\theta}',y))]&\geq 
		E_{\vec{\theta}\sim \mu}[k(f( \vec{\theta},x),f(\vec{\theta},y))]
	\end{align}
	
	Since the joint distribution $\gamma\in \mathcal{J}_{IND}(\mu,\mu)$ was arbitrary, 
	$$\inf_{ \gamma \in \mathcal{J}_{IND}(\mu,\mu)}E_{(\vec{\theta},\vec{\theta}')\sim \gamma}[k(f( \vec{\theta},x),f(\vec{\theta}',y))]\geq E_{\vec{\theta}\sim \mu}[k(f( \vec{\theta},x),f(\vec{\theta},y))]$$
	
	Further, since $\bigotimes_{n=1}^d \mathcal{L}(\theta_n,\theta_n)\in \mathcal{J}_{IND}(\mu,\mu)$, the inequality goes the other way,
	$$\inf_{ \gamma \in \mathcal{J}_{IND}(\mu,\mu)}E_{(\vec{\theta},\vec{\theta}')\sim \gamma}[k(f( \vec{\theta},x),f(\vec{\theta}',y))]\leq E_{\vec{\theta}\sim \mu}[k(f( \vec{\theta},x),f(\vec{\theta},y))]$$
	And so, equality ensues, 
	$\inf_{ \gamma \in \mathcal{J}_{IND}(\mu,\mu)}E_{(\vec{\theta},\vec{\theta}')\sim \gamma}[k(f( \vec{\theta},x),f(\vec{\theta}',y))]= E_{\vec{\theta}\sim \mu}[k(f( \vec{\theta},x),f(\vec{\theta},y))]$.
	Combining the above equations, we get that $
	\widetilde{W}_k(\mathcal{L}(X), \mathcal{L}(Y))=E[k(f(\vec{\theta},x),f(\vec{\theta},y))]$.
\end{proof}
	
	\subsection{Proof of Theorem \ref{thm:rdrp}}\label{pf:rdrp}
	\begin{proof}
		Denote the $i$th coordinate of $\vec{X}$ as $\vec{X}^{(i)}=f^{(i)}(\vec{\theta},\vec{x})$.
		\begin{align}\label{eq:wassrq}
			&\inf_{\gamma\in \mathcal{J}_{IND}(\mu,\mu)}E_{(\vec{\theta},\vec{\theta}')\sim \gamma}[||f(\vec{\theta},\vec{x})-f(\vec{\theta}',\vec{y})||_p^p]\\
			&=\inf_{\gamma\in \mathcal{J}_{IND}(\mu,\mu)} \sum_{i=1}^q E_{(\vec{\theta},\vec{\theta}')\sim \gamma}[|f^{(i)}(\vec{\theta},\vec{x})-f^{(i)}(\vec{\theta}',\vec{y})|^p]\\
			&\geq\sum_{i=1}^q \inf_{\gamma\in  \mathcal{J}_{IND}(\mu,\mu)}  E_{(\vec{\theta},\vec{\theta}')\sim \gamma}[|f^{(i)}(\vec{\theta},\vec{x})-f^{(i)}(\vec{\theta}',\vec{y})|^p]\\
			&=\sum_{i=1}^q  E_{\vec{\theta}\sim \mu}[|f^{(i)}(\vec{\theta},\vec{x})-f^{(i)}(\vec{\theta},\vec{y})|^p]&\text{by Equation \ref{eq:rdconc}}
		\end{align}
		The last equality is a result of Equation \ref{eq:rdconc} and since $h(x-y)=|x-y|^p$ is a quasi-antitone function. Since $\bigotimes_{n=1}^d \mathcal{L}(\theta_n,\theta_n)\in \mathcal{J}_{IND}(\mu,\mu)$,
		equality follows. Next,
		\begin{align*}
			\widetilde{W}_{||\cdot||_p,p}(\mathcal{L}(\vec{X}), \mathcal{L}(\vec{Y}))&=\left(\inf_{\gamma\in  \mathcal{J}_{IND}(\mu,\mu)}E_{(\vec{\theta},\vec{\theta}')\sim \gamma}[||f(\vec{\theta},\vec{x})-f(\vec{\theta}',\vec{y})||_p^p]\right)^{1/p}& \text{By Definition \ref{def:pwd} \& Section \ref{sec:background}}\\
			&= \left( \sum_{i=1}^qE_{\vec{\theta}\sim \mu}[|f^{(i)}(\vec{\theta},\vec{x})-f^{(i)}(\vec{\theta},\vec{y})|^p]\right)^{1/p}&\text{By eq. \ref{eq:wassrq}}\\
			&=\left( E_{\vec{\theta}\sim \mu}[||f(\vec{\theta},\vec{x})-f(\vec{\theta},\vec{y})||_p^p]\right)^{1/p}\\
		\end{align*}
	\end{proof}
	\newpage
	\subsection{Proof of Theorem \ref{thm:crnmc}}\label{pf:crnmc}
	
			\begin{proof}
		
		\underline{Part 1: coordinatewise monotone in $\Theta_n$:}
		Recall $\vec{X}\in \mathbb{R}^q$ and $\vec{X}^{(l)}=f^{(l)}(\vec{\theta}, \vec{x})$.
		Fix $j\in \{1, \ldots, d\}$. Define $\vec{\theta}_{1,-j}=\vec{\theta_1}\backslash \theta_{1,j}$. 
		
		\underline{Base step:}
		If $\theta_{1,j}\leq \theta_{1,j}'$ $\mu_j$-a.e., then for each $k\in \{1\ldots q\}$ and $\vec{\theta}_{1,-j}\sim\mu_{-j}$,
		$f^{(k)}((\vec{\theta}_{1,-j},\theta_{1,j}),\vec{x})\leq f^{(k)}((\vec{\theta}_{1,-j},\theta'_{1,j}),\vec{x})$ by monotonicity of $\vec{\theta}$.
		Thus for arbitrary $l\in \{1\ldots q\}$,
		\begin{align*}
			&f^{(l)}(\theta_2,f((\vec{\theta}_{1,-j},\theta_{1,j}),\vec{x}))\\
			&=f^{(l)}[\theta_2,(f^{(1)}((\vec{\theta}_{1,-j},\theta_{1,j}),\vec{x}), f^{(2)}((\vec{\theta}_{1,-j},\theta_{1,j}),\vec{x}),\ldots f^{(q)}((\vec{\theta}_{1,-j},\theta_{1,j}),\vec{x}))]\\
			&\leq f^{(l)}[\theta_2,(f^{(1)}((\vec{\theta}_{1,-j},\theta'_{1,j}),\vec{x}), f^{(2)}((\vec{\theta}_{1,-j},\theta_{1,j}),\vec{x}),\ldots f^{(q)}((\vec{\theta}_{1,-j},\theta_{1,j}),\vec{x}))]&\text{Since $f^{(l)}$ is monotone in $x_1$}\\
			&\leq f^{(l)}[\theta_2,(f^{(1)}((\vec{\theta}_{1,-j},\theta'_{1,j}),\vec{x}), f^{(2)}((\vec{\theta}_{1,-j},\theta'_{1,j}),\vec{x}),\ldots f^{(q)}((\vec{\theta}_{1,-j},\theta_{1,j}),\vec{x}))]&\text{Since $f^{(l)}$ is monotone in $x_2$}\\
			&\ldots\\
			&\leq f^{(l)}[\theta_2,(f^{(1)}((\vec{\theta}_{1,-j},\theta'_{1,j}),\vec{x}), f^{(2)}((\vec{\theta}_{1,-j},\theta'_{1,j}),\vec{x}),\ldots f^{(q)}((\vec{\theta}_{1,-j},\theta'_{1,j}),\vec{x}))]&\text{Since $f^{(l)}$ is monotone in $x_q$}\\
			&= f^{(l)}[\theta_2,f((\vec{\theta}_{1,-j},\theta'_{1,j}),
		\vec{x})]
		\end{align*}
		
		Since $j$ was arbitrary, $f(\vec{\theta_2},f(\vec{\theta}_1,\vec{x}))$, is coordinatewise monotone in $\vec{\theta}_1$. In other words, $f_2(\Theta_2,\vec{x})$ is monotone in $\Theta_2$ where $\Theta_2=(\vec{\theta}_1, \vec{\theta}_2)$.
		
		\underline{Induction step:}
		Next, suppose that  $f_n(\Theta_n,\vec{x})$ is monotone in $\Theta_n$. Show that $f_{n+1}(\Theta_{n+1},\vec{x})$ is monotone in $\Theta_{n+1}$
		
		Fix $i\in \{1\ldots,n\}$ and $j\in \{1\ldots, d\}$. Define $\theta_{j}$ as the $j$th coordinate of the $\Theta_n$. Define $\Theta_{n,-j}=\Theta_n\backslash \theta_{j}$. If $\theta_{j}\leq \theta_{j}'$ $\mu_i$-a.e., then for each $l\in \{1\ldots q\}$, $\Theta_{n,-j}\sim\mu_{-j}\times\mu^{n-1}$-a.e. and $\vec{x}$ $\pi$-a.e. the following inequality holds, because $f_{n}$ is coordinatewise monotone in $\Theta_n$ and $f$ is coordinatewise monotone in $\vec{\theta}$.
		\begin{align*}
			f^{(l)}_{n+1}(\Theta_{n+1},\vec{x})
			&=f^{(l)}_{n+1}((\Theta_{n+1,-j},\theta_{j}),\vec{x}))\\
			&=f^{(l)}(\vec{\theta}_{n+1},f_{n}((\Theta_{n,-j},\theta_{j}),\vec{x}))\\
			&\leq f^{(l)}(\vec{\theta}_{n+1},f_{n}((\Theta_{n,-j},\theta'_{j}),\vec{x}))
		\end{align*}  
		
		Since $j$ was arbitrary, $f_{n+1}(\Theta_{n+1},\vec{x})$, is coordinatewise monotone in $\Theta_{n+1}$ if $f_{n}(\Theta_{n},\vec{x})$, is coordinatewise monotone in $\Theta_{n}$. 
		
		By induction and base step, $f_n(\Theta_n,\vec{x})$ is coordinatewise monotone in $\Theta_n$, $n\geq 1$. 
		
		Theorem 4.7 of \cite{rudin1976principles} states that the composition of two continuous functions is continuous. Since $f(\vec{\theta},\vec{x})$ is continuous with respect to $\vec{\theta}$ and $\vec{x}$, $f_n(\Theta_n,\vec{x})$ is continuous with respect to $\Theta_n$. 
		Thus we apply Theorem \ref{prop:rd} if $x\in\mathbb{R}$ or Theorem \ref{thm:rdrp} if $\vec{x}\in\mathbb{R}^q$ to obtain Equation \ref{eq:crnmc}.
	\end{proof}

\subsection{Proof of Lemma \ref{lem:intflip}}\label{sec:intflip}
	\begin{proof}[Proof of Lemma \ref{lem:intflip}]\label{pf:intflip}
	Let $t\in \mathbb{R}$. We want to show that $P(f(g(U))\leq t)=P(f(U)\leq t)$
	
	\begin{align*}
		P(f(g(U))\leq t)&=\int_{[0,1]}I(f(g(u))\leq t) \d u\\
		&= \int_{a}^bI(f(a+b-u)\leq t) \d u + \int_{(a,b)^C}I(f(u)\leq t) \d u\\
		&= -\int_{b}^aI(f(v)\leq t) \d v + \int_{(a,b)^C}I(f(u)\leq t) \d u\\
		&= \int_{a}^bI(f(u)\leq t) \d u + \int_{(a,b)^C}I(f(u)\leq t) \d u\\
		&= \int_{[0,1]} I(f(u)\leq t) \d u \\
		&= P(f(U)\leq t)
	\end{align*}
\end{proof}

\subsection{Proof of Lemma \ref{lem:intmerge}}\label{sec:intmerge}
	\begin{proof}[Proof of Lemma \ref{lem:intmerge}]\label{pf:intmerge}
	Let $x\in \mathbb{R}$. We want to show that $P(f(h(U,\xi))\leq x)=P(f(U)\leq x)$.
	
	First note the following integrals.
	\begin{align}\label{eq:intmerge1}
		&\int_{(a,b)}I(f(h(u,1))\leq x) \d u\\
		&= \int_{(a,b)}I\left(f\left(\frac{u-a}{b-a}(t-a)+a\right)\leq x\right) \d u \\
		&= \int_{(a,t)}I\left(f\left(v\right)\leq x\right) \frac{b-a}{t-a}\d v &\text{by a change of variable $v=\frac{u-a}{b-a}(t-a)+a$}
	\end{align}
	
	\begin{align}\label{eq:intmerge2}
		&\int_{(a,b)}I(f(h(u,0))\leq x) \d u\\
		&= \int_{(a,b)}I\left(f\left(\frac{u-a}{b-a}(b-t)+t\right)\leq x\right) \d u \\
		&= \int_{(t,b)}I\left(f\left(v\right)\leq x\right) \frac{b-a}{b-t}\d v &\text{by a change of variable $v=\frac{u-a}{b-a}(b-t)+t$}
	\end{align}
	We show that $P(f(h(U,\xi))\leq x)=P(f(U)\leq x)$ as follows,
	\begin{align*}
		&P(f(h(U,\xi))\leq x)\\
		&= P(f(h(U,\xi))\leq x\mid \xi=1)P(\xi=1) + P(f(g(U,\xi))\leq x\mid \xi=0)P(\xi=0)\\
		&= P(f(h(U,1))\leq x)\frac{t-a}{b-a} + P(f(h(U,0))\leq x)\frac{b-t}{b-a}\\
		&= \left(\int_{(a,b)}I(f(h(u,1))\leq x) \d u + \int_{(a,b)^c}I(f(h(u,1))\leq x) \d u\right)\frac{t-a}{b-a} + \\ &\hspace{1cm}\left(\int_{(a,b)}I(f(h(u,0))\leq x) \d u + \int_{(a,b)^c}I(f(h(u,0))\leq x) \d u\right)\frac{b-t}{b-a}\\
		&= \int_{(a,b)}I(f(h(u,1))\leq x) \d u \frac{t-a}{b-a} + \int_{(a,b)}I(f(h(u,0))\leq x) \d u \frac{b-t}{b-a} + \\ &\hspace{1cm} \int_{(a,b)^c}I(f(u)\leq x) \d u\\
		&= \int_{(a,t)}I\left(f\left(v\right)\leq x\right) \frac{b-a}{t-a}\d v \frac{t-a}{b-a} + \int_{(t,b)}I\left(f\left(v\right)\leq x\right) \frac{b-a}{b-t}\d v \frac{b-t}{b-a} + \\ 
		&\hspace{1cm}
		\int_{(a,b)^c}I(f(u)\leq x) \d u  & \text{By eq. \ref{eq:intmerge1}, \ref{eq:intmerge2}}\\
		&= \int_{(a,t)}I\left(f\left(v\right)\leq x\right) \d v + \int_{(t,b)}I\left(f\left(v\right)\leq x\right) \d v + 
		\int_{(a,b)^c}I(f(u)\leq x) \d u\\
		&= \int_{[0,1]}I(f(u)\leq x) \d u\\
		&= P(f(U)\leq x)
	\end{align*}
\end{proof}

	\printbibliography

@misc{jacob,
author = {Pierre E. Jacob},
title = {Lecture notes for couplings and Monte Carlo},
url = {https://sites.google.com/site/pierrejacob/cmclectures?authuser=0}
}

@article{crn,
  title={Some guidelines and guarantees for common random numbers},
  author={Paul Glasserman and David D. Yao},
  journal={Management Science},
  year={1992},
  volume={38},
  pages={884-908},
  doi={10.1287/mnsc.38.6.884}
}

@article{crnconv,
  title={Studying convergence of Markov Chain Monte Carlo algorithms using coupled sample paths},
  author={Valen E. Johnson},
  journal={Journal of the American Statistical Association},
  year={1996},
  volume={91},
  pages={154-166},
  doi = {10.1080/01621459.1996.10476672}
}

@article{wasmeth,
author = {Qian Qin and James P. Hobert},
title = {Wasserstein-based methods for convergence complexity analysis of MCMC with applications},
volume = {32},
journal = {The Annals of Applied Probability},
number = {1},
publisher = {Institute of Mathematical Statistics},
pages = {124 -- 166},
year = {2022},
doi = {10.1214/21-AAP1673}
}

@article{waslinearmixedmodels,
title = {Dimension free convergence rates for Gibbs samplers for Bayesian linear mixed models},
journal = {Stochastic Processes and their Applications},
volume = {148},
pages = {25-67},
year = {2022},
doi = {10.1016/j.spa.2022.02.003},
author = {Zhumengmeng Jin and James P. Hobert}
}

@article{wasandoneshot,
	 author = {Neal Madras and Denise Sezer},
	 journal = {Bernoulli},
	 number = {3},
	 pages = {882--908},
	 title = {Quantitative bounds for Markov chain convergence: Wasserstein and total variation distances},
	 volume = {16},
	 year = {2010},
	 doi = {10.3150/09-bej238}
}

@article{iteratedranfunc,
	author = {Persi Diaconis and David Freedman},
	title = {Iterated random functions},
	journal = {SIAM Review},
	volume = {41},
	number = {1},
	pages = {45-76},
	year = {1999},
	doi = {10.1137/S0036144598338446}
}

@article{gibbs,
	 author = {Alison L. Gibbs and Francis Edward Su},
	 journal = {International Statistical Review / Revue Internationale de Statistique},
	 number = {3},
	 pages = {419--435},
	 publisher = {[Wiley, International Statistical Institute (ISI)]},
	 title = {On Choosing and Bounding Probability Metrics},
	 volume = {70},
	 year = {2002},
	 doi = {10.2307/1403865}
}

@article{gibbswass,
	author = {Alison L. Gibbs},
	title = {Convergence in the Wasserstein Metric for Markov Chain Monte Carlo Algorithms with Applications to Image Restoration},
	journal = {Stochastic Models},
	volume = {20},
	number = {4},
	pages = {473-492},
	year  = {2004},
	publisher = {Taylor & Francis},
	doi = {10.1081/STM-200033117}
}

@book{durrett,
  author = {Rick Durrett},
  edition = {4},
  publisher = {Cambridge Series in Statistical and Probabilistic Mathematics},
  title = {Probability: Theory and Examples},
  address = {New York},
  year = {2010},
  doi = {10.1017/9781108591034}
}

@article{honestexp,
	author = {James P. Hobert and Galin L. Jones},
	title = {{Honest Exploration of Intractable Probability Distributions via Markov Chain Monte Carlo}},
	volume = {16},
	journal = {Statistical Science},
	number = {4},
	publisher = {Institute of Mathematical Statistics},
	pages = {312 -- 334},
	year = {2001},
	doi = {10.1214/ss/1015346317}
}

@article{avcont,
title = {A Survey of Average Contractive Iterated Function Systems},
journal = {Journal of Difference Equations and Applications},
volume = {18},
number = {8},
pages = {1355-1380},
year = {2012},
doi = {10.1080/10236198.2011.610793},
author = {Örjan Stenflo}
}

@article{localcont,
title = {Locally Contractive Iterated Function Systems},
journal = {Annals of Probability},
volume = {27},
number = {4},
pages = {1952-1979},
year = {1999},
doi = {10.1214/aop/1022874823},
author = {David Steinsaltz}
}

@article{ifssurvey,
author = {Ramen Ghosh and Jakub Marecek},
year = {2022},
month = {11},
title = {Iterated Function Systems: A Comprehensive Survey},
eprint={2211.14661},
      archivePrefix={arXiv},
      primaryClass={math.PR}
}

@article{ifssurvey1,
author = {Krzysztof Leśniak and Nina Snigireva and Filip Strobin},
year = {2020},
journal = {Journal of Difference Equations and Applications},
volume = {26},
number = {8},
pages = {1114-1173},
title = {Weakly contractive iterated function systems and beyond: a manual},
doi = {10.1080/10236198.2020.1760258}
}

@article{cambanis1976,
author = {Stamatis Cambanis and Gordon Simons and William Stout},
title = {Inequalities for Ek(X, Y) when the marginals are fixed},
journal = {Zeitschrift für Wahrscheinlichkeitstheorie und Verwandte Gebiete},
volume = {36},
number = {4},
pages = {285-294},
year  = {1976},
doi = {10.1007/BF00532695},
}

@book{santambrogio,
  author = {Filippo Santambrogio},
  title = {Optimal Transport for Applied Mathematicians. Calculus of Variations, PDEs and Modeling},
  year = 2015,
  publisher = {Birkauser},
   edition = {1},
   doi = {10.1007/978-3-319-20828-2}
}

@article{biswas,
author = {Niloy Biswas and Lester Mackey},
year = {2021},
month = {12},
title = {Bounding Wasserstein distance with couplings},
doi = {arXiv:2112.03152},
primaryClass={stat.CO}
}

@book{handbook,
  editor={Steve Brooks and Andrew Gelman and Galin L. Jones and Xiao-Li Meng},
  title={Handbook of {M}arkov chain Monte Carlo},
  publisher={Chapman \& Hall},
  place={New York},
  year={2011},
  doi = {10.1201/b10905}
}

@article{probsurv,
  title={General state space {M}arkov chains and {MCMC} algorithms},
  author={Gareth O. Roberts and Jeffrey S. Rosenthal},
  journal={Probability surveys},
  volume={1},
  pages={20--71},
  year={2004},
  DOI={10.1214/154957804100000024},
  publisher={Institute of Mathematical Statistics and Bernoulli Society}
}

@book{firstlook,
   author = {Jeffrey S. Rosenthal},
   year = {2016},
   title = {A First Look at Rigorous Probability Theory},
   publisher = {World Scientific},
   edition = {2},
   doi = {10.1142/6300}
}

@book{billingsley,
   author = {Patrick Billingsley},
   year = {2012},
   title = {Probability and Measure: Anniversary Edition},
   publisher = {Wiley Series in Probability and Statistics},
   address = {New York},
   ISBN = {978-1-118-34191-9}
}

@book{villani,
   author = {Cédric Villani},
   year = {2008},
   title = {Optimal Transport: Old and New},
   publisher = {Springer Berlin Heidelberg},
   edition = {1},
   doi = {10.1007/978-3-540-71050-9}
}

@article{browniansynch,
author = {Krzysztof Burdzy and Zhen-Qing Chen and Peter Jones},
title = {{Synchronous couplings of reflected Brownian motions in smooth domains}},
volume = {50},
journal = {Illinois Journal of Mathematics},
number = {1-4},
publisher = {Duke University Press},
pages = {189 -- 268},
year = {2006},
doi = {10.1215/ijm/1258059475}
}

@misc{independentmetropolishastings,
      title={On importance sampling and independent Metropolis-Hastings with an unbounded weight function}, 
      author={George Deligiannidis and Pierre E. Jacob and El Mahdi Khribch and Guanyang Wang},
      year={2024},
      eprint={2411.09514},
      archivePrefix={arXiv},
      primaryClass={math.ST}
}

@article{sdesynch,
      title={Adapted Wasserstein distance between the laws of SDEs},
   volume={189},
   DOI={10.1016/j.spa.2025.104689},
   journal={Stochastic Processes and their Applications},
   publisher={Elsevier BV},
   author={Backhoff-Veraguas, Julio and Källblad, Sigrid and Robinson, Benjamin A.},
   year={2025},
   month=Nov, pages={104689}
}

@article{sixtacrn,
	author = {Sabrina Sixta and Jeffrey S. Rosenthal and Austin Brown},
	title = {Estimating MCMC convergence rates using common random number simulations},
	year={2026},
      journal={Stochastic Models},
      publisher={Informa UK Limited},
      archivePrefix={arXiv},
      DOI={10.1080/15326349.2026.2688991},
month=July, pages={1–42}
}

@article{cuturi,
  author={Peyré, Gabriel and Cuturi, Marco},
  title={Computational Optimal Transport: With Applications to Data Science},
  journal={Foundations and Trends in Machine Learning},
  year={2019},
  volume={11},
  number={5-6},
  pages={355-607},
  doi={10.1561/2200000073}}

@article{wassdnm,
	author = {Qian Qin and James P. Hobert},
	title = {Geometric convergence bounds for Markov chains in Wasserstein distance based on generalized drift and contraction conditions},
	volume = {58},
journal = {Annales de l'Institut Henri Poincaré, Probabilités et Statistiques},
number = {2},
publisher = {Institut Henri Poincaré},
pages = {872--889},
year = {2022},
doi = {10.1214/21-AIHP1195}
}

@article{brenier,
	author = {Yann Brenier},
	title = {Polar factorization and monotone rearrangement of vector-valued functions},
	volume = {44},
journal = {Communications on Pure and Applied Mathematics},
number = {4},
pages = {375--417},
year = {1991},
doi = {10.1002/cpa.3160440402}
}

@BOOK{inequalities,
  author={Hardy, G. H. and Littlewood, J. E. and Pólya, G.},
  title={Inequalities},
  year={1934},
  volume={},
  number={},
  edition = {1},
  publisher = {Cambridge: At the University Press}
}

@article{rearrangefunc,
title = {Rearrangements of functions},
journal = {Journal of Functional Analysis},
volume = {66},
number = {3},
pages = {432-438},
year = {1986},
doi = {10.1016/0022-1236(86)90067-4},
author = {J.A. Crowe and J.A. Zweibel and P.C. Rosenbloom}
}

@article{quantile1, 
   volume={78},
   title = {Quantile and Probability Curves Without Crossing},
   author = {Victor Chernozhukov and Iván Fernández-Val and Alfred Galichon},
   doi={10.3982/ecta7880},
   number={3},
   journal={Econometrica},
   publisher={The Econometric Society},
   year={2010},
   pages={1093–1125} }

@article{equimeasurable,
 author = {Binyamin Schwarz},
 journal = {Journal of Mathematics and Mechanics},
 publisher = {Indiana University Mathematics Journal},
 number = {3},
 pages = {401--422},
 title = {On the Extrema of the Frequencies of Nonhomogeneous Strings with Equimeasurable Density},
 volume = {10},
 year = {1961}
}

@article{equimeasurable2,
  author = {Philip Savoye},
 journal = {The Rocky Mountain Journal of Mathematics},
 number = {1},
 pages = {281--293},
 publisher = {Rocky Mountain Mathematics Consortium},
 title = {Equimeasurable Rearragnements of Functions and Fourth Order Boundary Value Problems},
 volume = {26},
 year = {1996},
DOI = {10.1216/rmjm/1181072116}
}

@BOOK{Bog07,
  title = {Measure theory. {V}ol. {I}, {II}},
  publisher = {Springer-Verlag},
  year = {2007},
  author = {Bogachev, Vladimir I.},
  address = {Berlin},
  doi = {10.1007/978-3-540-34514-5}
}

@book{bridgeford2025measure,
  author    = {Eric W. Bridgeford},
  title     = {Measure Theoretic Probability Theory: A Hands-On, Proof-Based Approach},
  year      = {2025},
  url       = {https://ericwb.me/prob_book/coverpage.html}
}

@article{Sagiv2019Wasserstein,
  author  = {Sagiv, Amir},
  title   = {The Wasserstein Distances Between Pushed-Forward Measures
             with Applications to Uncertainty Quantification},
  journal = {Communications in Mathematical Sciences},
 number = {3},
 pages = {707--724},
 publisher = {International Press},
 volume = {18},
 year = {2020},
DOI = {10.4310/CMS.2020.v18.n3.a6}
}

@article{Carlier2025Quantitative,
  author  = {Carlier, Guillaume and Delalande, Alex and M{\'e}rigot, Quentin},
  title   = {Quantitative Stability of the Pushforward Operation by an Optimal Transport Map},
  journal = {Foundations of Computational Mathematics},
  volume  = {25},
  number  = {4},
  pages   = {1259--1286},
  year    = {2025},
  doi     = {10.1007/s10208-024-09669-4}
}

@book{vaart,
  author    = {A. W. van der Vaart},
  title     = {Asymptotic Statistics},
  year={1998},
  edition = {1},
  series={Cambridge Series in Statistical and Probabilistic Mathematics},
  publisher = {Cambridge University Press},
  doi = {10.1017/CBO9780511802256}
}

@book{rudin1976principles,
  title={Principles of Mathematical Analysis},
  author={Rudin, Walter},
  edition={3rd},
  year={1976},
  publisher={McGraw-Hill},
  address={New York},
  isbn={978-0070542358}
}
\end{document}